\documentclass[aps,prb,preprint,superscriptaddress,reprint,preprintnumbers,
amssymb]{revtex4-1}

\usepackage{graphicx}
\usepackage[latin1]{inputenc}
\usepackage{braket}

\usepackage{units}
\usepackage{wrapfig}
\usepackage{graphicx}
\usepackage{amsmath}
\usepackage{rotating}
\usepackage{color}
\usepackage{amsfonts}
\usepackage{amssymb}
\usepackage{moreverb}
\usepackage{url}
\usepackage{epsf}
\usepackage{multirow}
\usepackage{subfigure}
\usepackage{morefloats}
\usepackage{threeparttable}
\usepackage{cancel}
\usepackage{bm}
\usepackage[version=3]{mhchem}

\begin{document}

\title{
Fragment Approach to Constrained Density Functional Theory Calculations using Daubechies Wavelets
}
%\title{
%

%}

\author{Laura E.\ Ratcliff}
\email{lratcliff@anl.gov}
\affiliation{Argonne Leadership Computing Facility, Argonne National Laboratory, Illinois 60439, USA}
\affiliation{Univ.\ Grenoble Alpes, CEA, INAC-SP2M, L\_Sim, F-38000, Grenoble, France}
%\affiliation{CEA, INAC-SP2M, L\_Sim, F-38000 Grenoble, France}
%\altaffiliation{Now at Argonne Leadership Computing Facility, Argonne National Laboratory, Illinois 60439, USA}

\author{Luigi Genovese}
\affiliation{Univ.\ Grenoble Alpes, CEA, INAC-SP2M, L\_Sim, F-38000, Grenoble, France}
%\affiliation{CEA, INAC-SP2M, L\_Sim, F-38000 Grenoble, France}

\author{Stephan Mohr}
\affiliation{Univ.\ Grenoble Alpes, CEA, INAC-SP2M, L\_Sim, F-38000, Grenoble, France}
%\affiliation{CEA, INAC-SP2M, L\_Sim, F-38000 Grenoble, France}

\author{Thierry Deutsch}
\affiliation{Univ.\ Grenoble Alpes, CEA, INAC-SP2M, L\_Sim, F-38000, Grenoble, France}
%\affiliation{CEA, INAC-SP2M, L\_Sim, F-38000 Grenoble, France}

\date{\today}

\begin{abstract}
In a recent paper we presented a linear scaling Kohn-Sham density functional theory (DFT) code based on Daubechies wavelets,
where a minimal set of localized support functions is optimized \emph{in situ} and therefore adapted to the chemical properties
of the molecular system.
Thanks to the systematically controllable accuracy of the underlying basis set, this approach is able to provide an optimal contracted basis for a 
given system: accuracies for ground state energies and atomic forces are of the same quality as an uncontracted, cubic scaling approach.
This basis set offers, by construction, a natural subset where the density matrix of the system can be projected.
In this paper we demonstrate the flexibility of this minimal basis formalism in providing a basis set that can be reused \emph{as-is}, i.e.\ without reoptimization, for charge-constrained DFT calculations within a \emph{fragment} approach. Support functions, represented in the underlying wavelet grid, of the template fragments are roto-translated with high numerical precision to the required positions and used as projectors for the charge weight function. 
We demonstrate the interest of this approach to express highly precise and efficient calculations for preparing diabatic states and for the computational setup of systems in complex environments.
\end{abstract}

\maketitle

\section{Introduction}

Density functional theory (DFT)~\cite{hohenberg42,kohn43} is arguably the most 
popular approach to electronic structure calculations for a wide range of 
systems.  However, it suffers from various well-known limitations, like for 
example the self-interaction problem~\cite{PhysRevB.23.5048,1.476859} which can 
result in electron delocalization errors, and the fact that it is in principle a 
ground state theory only.  For these reasons, the DFT formalism has been extended in 
the form of constrained DFT (CDFT)~\cite{PhysRevLett.53.2512} to include an 
additional constraint on the density, so that the lowest energy state 
\emph{satisfying a given condition} can instead be found. When a reasonable 
guess for such a condition is at hand, it can therefore be used both to find a 
particular excited state of the system and to localize the electronic density in 
such a way as to prevent spurious delocalization, and thus provides a way of 
overcoming the above problems.  Of course, time-dependent DFT 
(TDDFT)~\cite{tddft} can be used to find multiple excited states, however when 
one is interested in a particular excited state CDFT can be advantageous, 
especially given that the additional costs associated with adding a constraint 
are relatively low.  Furthermore, TDDFT also suffers from self interaction 
problems, and can give inaccurate results for certain types of excited states, 
including charge transfer excitations.

Constrained DFT has been implemented in a number of codes, using both localized 
basis sets~\cite{conquest_cdft} and plane 
waves~\cite{oberhofer:064101,doi:10.1021/ct200570u}, and has been successfully 
applied in a variety of contexts, including charge constrained molecular 
dynamics~\cite{oberhofer:064101}, the calculation of the correct energy 
alignment of metal/molecule interfaces~\cite{PhysRevB.88.165112} and the 
calculation of electronic coupling matrix 
elements~\cite{wu:164105,oberhofer:244105}.  For a general overview of CDFT see 
Ref.~\onlinecite{doi:10.1021/cr200148b}.

As with all DFT calculations, the choice of basis set has a large impact on both 
the accuracy and computational cost of CDFT.  One way of accessing large systems 
is to reformulate the standard cubic scaling approach to DFT in terms of 
localized orbitals, or `support functions', which we will discuss further in the 
following section.  We wish to perform CDFT calculations on large systems using 
such an approach, whilst maintaining the high accuracy associated with 
systematic basis sets.  As such, we require a basis set which is at the same 
time localized and systematic.  For this reason, we have chosen to use a 
Daubechies wavelet basis set~\cite{Daubechies}, as it is a systematic basis set 
exhibiting the desired properties of compact support in both real and Fourier 
space and can be chosen to be orthogonal.  Wavelet basis sets have 
an inherent flexibility, in that they allow for multiresolution grids, which is 
particularly useful for inhomogenous systems.  Combined with the ability to 
explicitly treat charged systems in open boundary conditions, wavelets provide 
an ideal basis set for accurate CDFT calculations of large systems.  

In this paper, we show that the combination of a support function approach with a wavelet basis 
set allows for the definition of a flexible fragment based approach to CDFT, 
which can further reduce the computational cost, particularly for very large 
systems.  In this approach, a set of support functions are optimized for an 
isolated (small) molecule, or `fragment', and reused as a fixed basis in a 
larger system containing many of these molecules e.g.\ a solvent.  It is then 
straightforward to associate the constrained charge with a given fragment using 
a L\"owdin like definition of the CDFT weight function.  However, in the larger 
system each molecule may well have a different orientation and so the support 
functions, which are described in terms of the fixed wavelet grid, cannot simply 
be duplicated for each molecule.  

Therefore, we have developed a scheme to 
reformat the support functions for arbitrary roto-translations using 
interpolating scaling functions.  This interpolation,
thanks to the properties of the underlying basis set,
results in only a negligible loss of 
accuracy and so the support functions can be directly reused, reducing the 
computational cost by an order of magnitude compared to optimizing the support 
functions from scratch for the full system.

%\luigi{Here in my opinion a stronger motivation for the paper should be 
%written. In other terms, why wavelets, why this approach? I think that there are 
%good elements in the next sections which might be anticipated here}.
In the following sections we will first summarize our approach to large scale DFT 
calculations using localized support functions represented in a wavelet basis 
set~\cite{linear_paper}, as implemented in the BigDFT electronic structure 
code~\cite{genovese:014109}.  We will then outline our 
implementation of CDFT, following which we will explain our fragment approach, 
including a description of the reformatting scheme,  validating our method with calculations on 
prototypical systems.  Finally, we will present an application of 
CDFT for the fullerene $\textrm{C}_{60}$ in two different environments, through which 
we will demonstrate the flexibility and potential of a fragment based approach.

\section{Methodology}

\subsection{Linear scaling DFT with wavelets}

\begin{figure}
\includegraphics[scale=0.35]{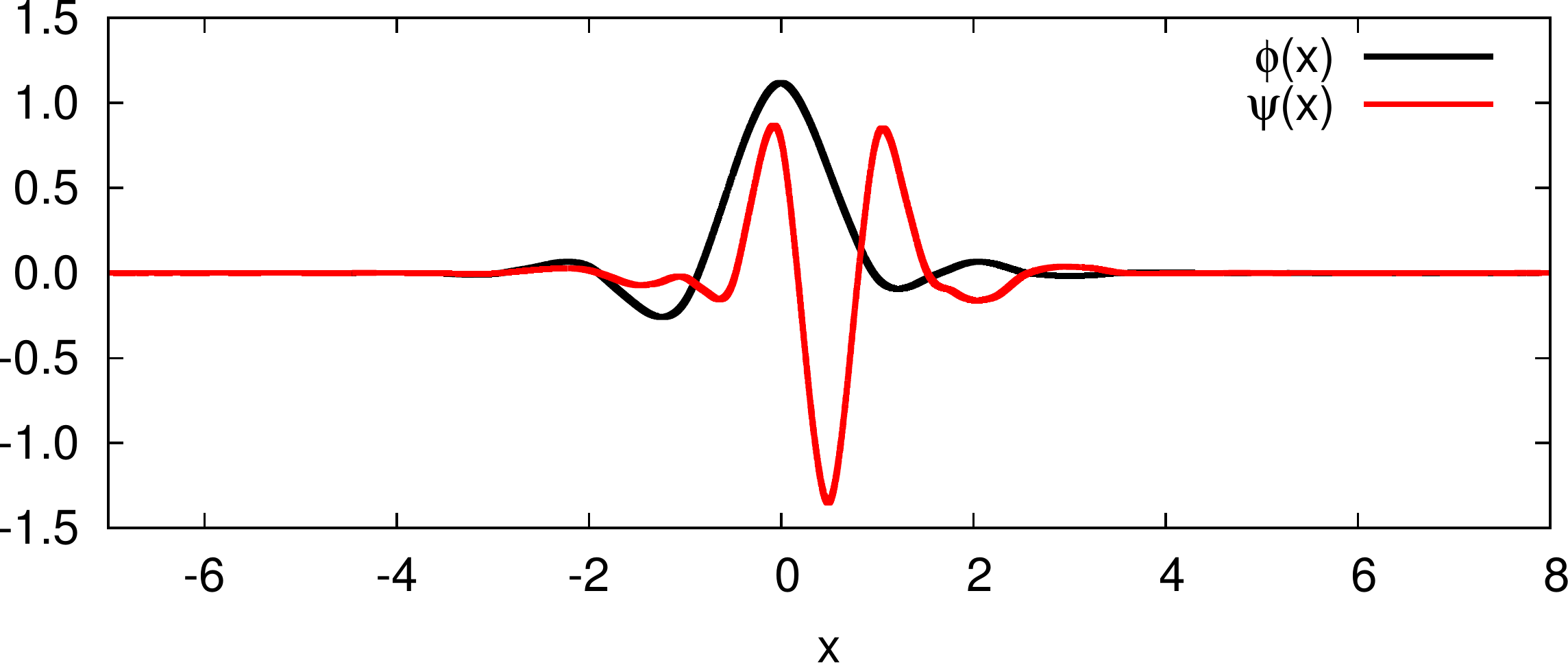}
\caption{Least asymmetric Daubechies wavelet family of order $2m=16$; both the 
scaling function $\phi(x)$ and
wavelet $\psi(x)$ differ from zero only within the interval 
$[1-m,m]$.\label{fig:Daub_16}}
\end{figure}

We and others have recently presented a newly developed method for DFT 
calculations on large systems, which combines the use of a minimal localized 
basis of `support functions' with the use of an underlying wavelet basis 
set~\cite{linear_paper}.  This method has been implemented in BigDFT, which uses 
the orthogonal least asymmetric Daubechies family of order 16, which are 
depicted in Fig.~\ref{fig:Daub_16}.  The Kohn-Sham (KS) orbitals are expressed 
in terms of the support functions via a set of coefficients $c_i^\alpha$:
\begin{eqnarray}
 \Ket{\varPsi_i}=\sum_\alpha c_i^\alpha\Ket{\phi_\alpha},
 \label{eq:expansion}
\end{eqnarray}
where the support functions are represented directly in the wavelet basis set localized on a 3 dimensional grid, 
so that they can be thought of as adaptively contracted wavelets.  Rather than 
working directly with the KS orbitals, we instead work in terms of the density 
matrix, $\rho(\mathbf{r},\mathbf{r}')$, which is itself defined in terms of the 
support functions and the density kernel, $K^{\alpha\beta}$:
\begin{eqnarray} 
 \rho(\mathbf{r},\mathbf{r}')=\sum_{\alpha,\beta} 
\phi_\alpha(\mathbf{r})K^{\alpha\beta}\phi_\beta(\mathbf{r'}).
 \label{eq:density}
\end{eqnarray}
The density matrix has been shown to decay exponentially with distance for 
systems with a gap thanks to the so-called nearsightedness 
principle~\cite{kohn16,Prodan88,ismail46,kohn45,he81,baer1997sparsity}, and thus 
a formulation in terms of the density matrix allows us to take advantage of this 
to achieve linear scaling with the number of atoms in the system, thereby avoiding 
the cubic scaling of standard approaches to DFT. 
From this the charge density is calculated directly from the support functions 
and density kernel.  Similarly, the band structure energy and the charge of the system can be 
calculated from the density kernel via:
\begin{eqnarray}
E_{\mathrm{BS}}=\mathrm{Tr}\left[\mathbf{K}\mathbf{H}\right],
\quad 
N = \mathrm{Tr}\left[\mathbf{K}\mathbf{S}\right]
\end{eqnarray}
where $\mathbf{H}$ indicates the Hamiltonian matrix in the basis of the support 
functions, and $\mathbf S$ is the support function overlap matrix.

The support function formalism allows one to \emph{map} the degrees of freedom of KS orbitals into a \emph{localized} description,
that can be directly put in relation with atomic positions.
 In practice, the support functions are truncated within spherical localization 
regions with a user-defined radius, and some additional truncation must be 
applied to the density kernel which is then exploited via sparse matrix algebra 
to achieve a fully linear scaling algorithm.
Therefore, to some extent, a given support function $\phi_\alpha$ can be associated to the atom $a$ where its localization region is centered.
In order to achieve accurate 
results, both the support functions and density kernel are optimized during the 
calculation, that is the energy is minimized with respect to both quantities.  
Providing the localization regions are sufficiently large, this results in a 
minimal localized basis set with an accuracy equivalent to the underlying basis 
set.

The general scheme is common to other basis optimization for density-matrix 
minimization based linear scaling DFT codes, e.g.\ ONETEP~\cite{skylaris:084119} 
and Conquest~\cite{0953-8984-22-7-074207}, with the addition of a few novel 
features.  These include the application of a confining potential to the KS 
Hamiltonian, which ensures the support functions remain localized.  In order to 
apply the confining potential consistently, we also enforce an approximate 
orthogonality constraint on the support functions, in contrast with other 
approaches which use fully non-orthogonal support 
functions~\cite{skylaris:084119,0953-8984-22-7-074207}.  Furthermore, the 
properties of the description in the wavelet basis are such that the algorithm guarantees that the Pulay contribution to the atomic forces can safely be neglected, so the forces can be calculated accurately and cheaply.

The method can be divided into two key components: the optimization of the 
support functions (either with or without a confining potential), and the 
optimization of the density kernel.  This latter point can be achieved via a 
choice of schemes, as detailed previously~\cite{linear_paper}.  These include a 
direct minimization approach, where the coefficients $c_i^\alpha$ are first 
updated using DIIS or steepest descents to minimize the band structure energy and 
then used to construct the density kernel, and the Fermi Operator Expansion 
(FOE) method, where the density kernel is expressed as a function of the 
Hamiltonian matrix which can be evaluated numerically using a Chebyshev 
polynomial expansion. Close attention has also been paid to the parallelization of the code, 
such that massively parallel machines can be exploited to perform large scale 
calculations (see also Ref.\cite{universal}). For charged calculations, we have found the use of the 
direct minimization method to be the most suitable, due to a reduction in the 
occurrence of charge sloshing during convergence, and the flexibility afforded 
by working with the wavefunction coefficients rather than directly with the 
density kernel as in the FOE method.
%  In principle the support function optimization could be omitted 
%so that the calculation would instead be performed with a fixed localized basis 
%set, but this would of course reduce the accuracy.

\subsection{Atomic charge analysis}
The mapping between electronic and localized degrees of freedom which is provided by the support function formalism allows one to perform an accurate atomic charge analysis, 
meaning that each atom is assigned a partial net charge, such that the electrostatic properties of the system are conserved. 
Obviously this conservation is only possible within a certain limit, 
as one is mapping a continuous quantity (the electronic charge) to a discrete quantity (the atomic point charges).
If, however, the error introduced by this mapping onto point charges is small enough, the system under investigation can 
be reasonably approximated by a simple setup of charged point particles, which paves the way for future applications such as coupling different levels of accuracy within the same calculation.

Given the overlap matrix $\mathbf{S}$ and the density kernel $\mathbf{K}$, the partial charge located on atom $a$ can be defined by the so-called L\"{o}wdin charge:
\begin{equation}
  q_a = \sum_{\alpha'}^{(a)} \left(\mathbf{S}^{1/2}\mathbf{K}\mathbf{S}^{1/2}\right)_{\alpha'\alpha'},
\end{equation}
where the sum runs over all support functions $\alpha'$ which are located on atom $a$.
Obviously $\sum_a q_a = \mathrm{tr}(\mathbf{K}\mathbf{S}) = N$, i.e.\ the total charge (the monopole) is conserved.
In order to check whether higher multipoles can also be conserved, we compared the dipole moment calculated using this approach with that
calculated using the continuous charge density. In addition a comparison with the dipole moment calculated with the cubic scaling version of BigDFT was done as a reference. The values for a strongly polarized molecule (\ce{H2O}) and a non-polarized one (\ce{C60}) are given in Tab.~\ref{tab:dipole_moments}.\\
\begin{table*}
 \centering
 \begin{tabular}{l | r | r r}
 \hline\hline
  & \multicolumn{1}{c}{cubic} & \multicolumn{2}{c}{linear} \\
    \cline{2-2} \cline{3-4}
  & \multicolumn{1}{c}{exact dipole}  & \multicolumn{1}{c}{exact dipole} &
    \multicolumn{1}{c}{point charge approx.} \\
  \hline
  \ce{H2O} & (0.463, -0.506, -0.186)    & (0.466, -0.510, -0.187)   & (0.606, -0.668, -0.247) \\
    norm & 0.711 & 0.716 & 0.935 \\
  $d_{ex} \cdot d$	 & 1 & 0.9999996 & 0.9999897\\
  \hline
  \ce{C60} & (-0.0004, -0.0004, -0.0004) & (-0.025, -0.025, -0.025) & (-0.055, -0.055, -0.055) \\
 \hline
 \end{tabular}
 \caption{Dipole moments calculated using the exact charge density for the cubic and linear scaling approaches, respectively, and using the partial atomic point charges. All values are given in atomic units.}
 \label{tab:dipole_moments}
\end{table*}
As a second test of the reliability of our method we directly compared the atomic point charges with those calculated by performing a Bader charge analysis of the charge density calculated using the cubic scaling approach. As can be seen from Tab.~\ref{tab:atomic_charges}, the differences between the exact results are smaller with our approach. In particular, for the \ce{C60} fullerene, reasons of symmetry impose that the charge should be equally distributed and no atom should carry a net charge. As can be seen, the L\"{o}wdin procedure comes closer to this result than the Bader analysis.
\begin{table}
 \centering
 \begin{tabular}{l r r r}
 \hline\hline
  & \multicolumn{1}{c}{cubic -- Bader} & & \multicolumn{1}{c}{linear -- L\"{o}wdin} \\
    \hline
  \ce{H2O} & (-1.27, 0.61, 0.67) & & (-0.83, 0.42, 0.41) \\
  \ce{C60} & $0.061 \pm 0.045$  & & $0.003 \pm 0.002$    \\
 \hline\hline
 \end{tabular}
 \caption{Atomic point charges, calculated by a Bader analysis of the charge density from the  cubic scaling approach, and the L\"{o}wdin procedure using the density kernel and overlap matrix from the linear scaling approach. For \ce{H2O} we indicate the values on all three atoms, for \ce{C60} we give the mean of the absolute values together with the standard deviation. All values are given in atomic units.}
 \label{tab:atomic_charges}
\end{table}

\subsection{Constrained DFT}\label{sec:cdft}

The general idea of constrained DFT is to force a charge to remain localized in a given 
region of the simulation space. This is achieved via the addition of a Lagrange 
multiplier term to the Kohn-Sham energy functional which enforces a given 
constraint on the resulting electronic density, so that rather than being the 
ground-state density of the system, the density instead corresponds to a 
particular excited state.  This Lagrange multiplier can also be thought of as an 
additional applied potential, otherwise referred to as the constraining 
potential.  The constraint can also take a number of other forms, but for the purposes of 
this work we are interested only in constraining the charge.  The new functional 
therefore becomes:
\begin{eqnarray}\label{eq:consfunc}
W\left[\rho,V_c\right]=E_{\mathrm{KS}}\left[\rho\right]+V_c\left(\int 
w_c\left(\mathbf{r}\right)\rho\left(\mathbf{r}\right)\mathrm{d}\mathbf{r}
-N_c\right),
\end{eqnarray}
where $E_{\mathrm{KS}}$ is the Kohn-Sham energy functional, $V_c$ is the aforementioned Lagrange multiplier, $N_c$ is the required 
charge within the specified region and $w_c\left(\mathbf{r}\right)$ is a weight 
function which defines this region.  The weight function and $N_c$ are defined 
in advance, however the value of $V_c$ which correctly enforces the constraint 
must be found during the calculation.  Wu and Van 
Voorhis~\cite{PhysRevA.72.024502} demonstrated that the lowest energy state for 
which the constraint is correctly applied is in fact a maximum with respect to 
$V_c$ and so it becomes possible to efficiently determine the correct $V_c$.  It 
is also straightforward to add multiple constraints to the system, and indeed 
one is frequently interested in constraining the charge difference between two 
regions. This feature is important for the simulation of charge-transfer excitations within the CDFT formalism.

Rewriting the new functional (Eq.~\ref{eq:consfunc}) in density matrix form 
as~\cite{conquest_cdft}:
\begin{eqnarray}\label{eq:consfunctr}
W\left[\rho,V_c\right]=E_{\mathrm{KS}}\left[\rho\right]+V_c\left(\mathrm{Tr}
\left[\mathbf{K}\mathbf{w}_c\right]-N_c\right),
\end{eqnarray}
the charge constraint is easily added to the existing algorithm in BigDFT.  This 
construction requires the weight matrix, $w_{\alpha\beta}$, which is defined via 
the weight function as
\begin{eqnarray}
w_{\alpha\beta}=\int\phi_\alpha\left(\mathbf{r}\right)w_c\left(\mathbf{r}
\right)\phi_\beta\left(\mathbf{r}\right)\textrm{d}\mathbf{r} .
\end{eqnarray}
It then remains to define the weight function, for which a number of different 
schemes exist.  The support function approach used here lends itself to a 
L\"{o}wdin like definition, which is analogous to that used above to determine atomic charges.
Using this approach we directly construct the weight matrix via
\begin{eqnarray}
w_{\alpha\beta}=\left(\mathbf{S}^{\frac{1}{2}}\mathbf{P}\mathbf{S}^{\frac{1}{2}}
\right)_{\alpha\beta},
\end{eqnarray}
where $\mathbf{S}$ is the overlap matrix between support functions and 
$\mathbf{P}$ is a projection matrix, defined as $1$ for all support functions 
belonging to the region where a constraint is being applied, and $0$ 
elsewhere.  Alternatively, if one is constraining a charge difference between 
two regions, it should be set to $1$ on one of the regions, $-1$ on the 
other, and $0$ elsewhere. 

The final step, once the weight function has been defined, is to derive a scheme 
for finding the correct value of $V_c$ for a given charge constraint value, 
$N_c$.  There are two possible approaches to the optimization.  
In the first approach, one can find the optimum value of $V_c$ at each step of 
the self-consistent density optimization, i.e.\ the ground state density is 
updated in an outer loop, with $V_c$ updated in an inner loop. 
Alternatively, the second approach consists of fully minimizing the functional 
$W\left[\rho,V_c\right]$ of Eq.~\ref{eq:consfunctr} with respect to the density 
for a fixed value of $V_c$, updating $V_c$ and finding the new minimum density, 
then repeating to convergence, i.e.\ the maximization with respect to $V_c$ is 
performed in an outer loop with the ground state density found self-consistently 
in an inner loop. We chose the latter, as it was observed to be more stable.  We 
use Newton's method to update $V_c$, with the second derivative calculated using 
a finite difference approach.

\subsection{Fragment approach}\label{sec:frag}

The combination of the novel features described above and the use of a wavelet 
basis set make this approach ideal for the application of CDFT to large systems. 
 In particular the ability to reuse the support functions can result in 
significant savings for e.g.\ geometry optimizations and calculations on charged 
systems, as previously demonstrated~\cite{linear_paper}.  Furthermore, this idea 
of support function reuse can be extended to a fragment based approach, which is 
similar to the so-called fragment orbital method which has been used to calculate 
electronic coupling matrix elements~\cite{oberhofer:244105,frag_adf1,frag_adf2}. 

\begin{figure}
\includegraphics[scale=0.12]{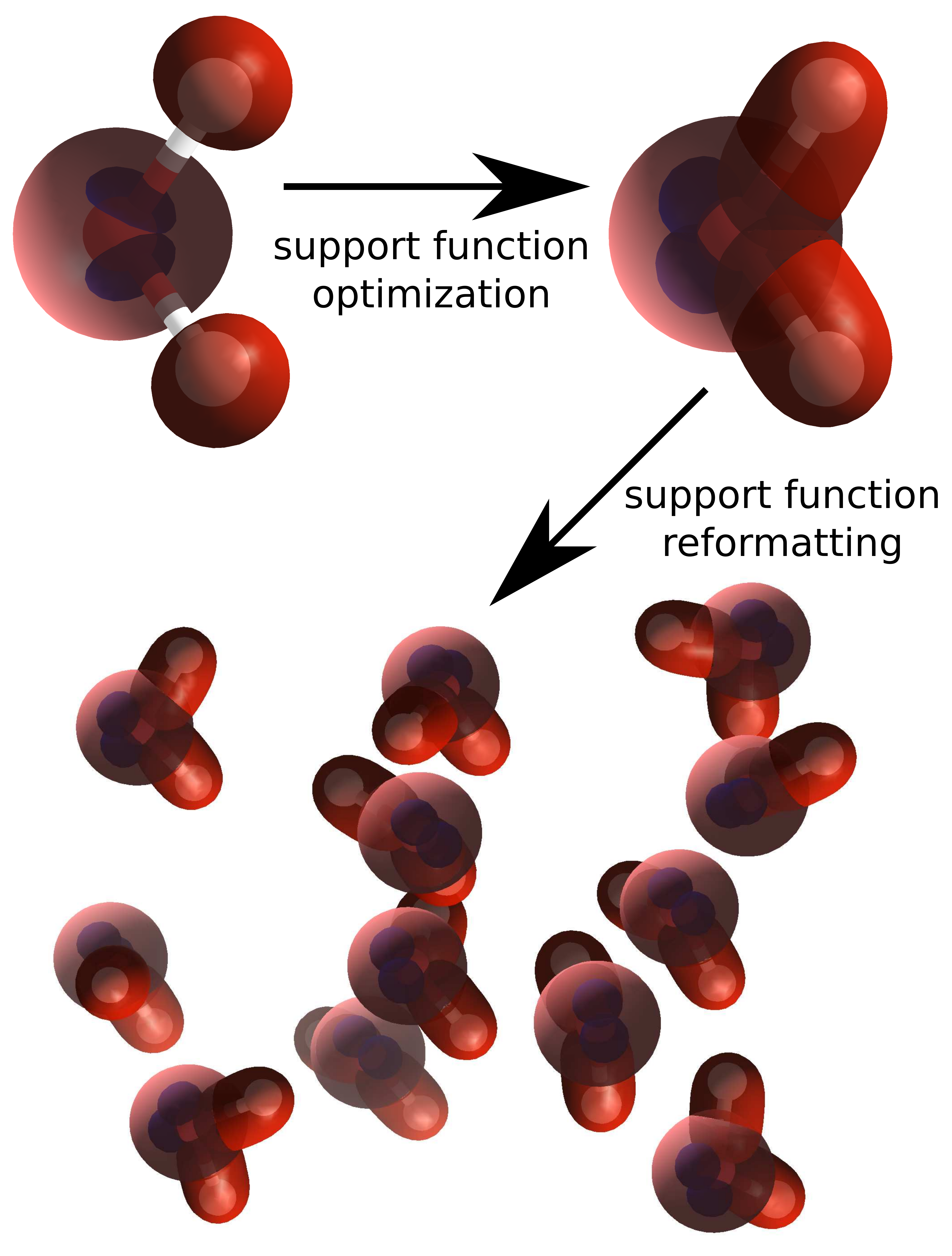}
\caption{The fragment approach as illustrated for a cluster of water molecules: 
the support functions are initially optimized for an isolated water molecule and 
then duplicated for a collection of water molecules, avoiding the need for 
optimization in the larger system.}\label{fig:fragment}
\end{figure}

The central idea is to take a group of atoms, or more specifically an isolated 
molecule, and fully optimize the support functions.  These support functions are 
then used as a fixed basis for a system containing several molecules, as 
illustrated in Fig.~\ref{fig:fragment} for a simple example.
We refer to the initial molecule for which the support functions were optimized 
as the `template' molecule.

As the support functions are kept fixed in the fragment approach, 
$w_{\alpha\beta}$ need only be calculated once at the start of the calculation, 
after which it remains fixed.  Furthermore, due to the quasi-orthogonality of 
the support functions, when the fragment approximation is justified $\mathbf{S}^{\frac{1}{2}}$ can in general be calculated 
using a Taylor approximation, and so the calculation of the weight matrix adds 
very little overhead to the calculation.  
%Where fragments are closer together 
%this quasi-orthogonality can break down; in such cases %it may not be possible to 
%use a Taylor approximation, but more generally for such %a case the validity 
%of the fragment approximation should also be questioned.

In this work we focus on systems where the respective fragments are well 
defined, and thus the support functions generated from the isolated fragments 
can be used for the full system with a minimal impact on the accuracy.  In cases 
where electrons are being added to a fragment, it is important to ensure that 
the lowest unoccupied molecular orbital (LUMO) and, if necessary, the next few 
states in energy are sufficiently well represented by the support function 
basis.  As discussed elsewhere~\cite{linear_paper}, this can be achieved using the direct 
minimization formalism to optimize a few additional states during the isolated 
calculation without adding a charge to the system.  For systems where the 
fragments are less well defined the implementation could in principle be 
extended to further optimize the support functions for the combined system, 
either in a neutral state or while the charge constraint is being enforced.  In 
such cases, the L\"{o}wdin approach is also expected to be less accurate, and so 
it would be desirable to use an alternative form for the weight function.

Finally, it should be mentioned that there are some subtleties related to the 
initial guess for the charge density.  This depends on the initial density 
kernel, which is constructed from the fragment KS orbitals.  For neutral 
calculations it is straightforward to use the fragment orbitals and occupancies 
directly from the isolated calculations, however for charged calculations some 
additional input is required.  One approach would be to occupy the fragment 
orbitals in order of their energies, however this can lead to charge 
distributions which are significantly different from the required constraint.  
This can result in slow 
convergence, or even worse, problems with local minima.  A better approach 
should therefore take into account the effect of the constraining potential on 
the fragment orbital energies.  This can be done by assigning occupation numbers 
so that any excess/deficit in charge is localized on the same fragment as the 
constraint, so that the initial density already satisfies the charge constraint. 
 Alternatively, the risk of encountering a local minimum can be reduced by 
adding a degree of noise to the fragment orbitals, or by completely randomizing 
the initial guess, subject to the correct overall charge.  However, such an 
approach is in general much slower to converge, and thus the latter strategy is 
generally not recommended.

\subsection{Reformatting scheme for roto-translations}

As the support functions are defined in terms of an underlying grid of wavelets, 
in order to implement a fragment approach it is necessary to have a scheme for 
reformatting the support functions due to a change in atomic positions.  
%Here, 
%we use an interpolation scheme, which we shall describe below.  For a rigid 
%shift in atomic positions, this procedure is relatively straightforward.  
We are frequently interested in situations where a molecule has been 
rotated and translated, for example when calculating electronic coupling matrix elements in a 
dimer for varying angles between the two monomers.  Therefore, we have developed and implemented a scheme for reformatting the support functions, given the axis and angle of rotation between some initial and final positions for a given fragment mass center.

A fragment of the system is defined by the user via a list of atomic 
positions, which should of course be in bijection with the atom list defining 
the template fragment. Therefore the first problem is to identify the 
combination of translation and rotation which sends the template fragment to the 
position of the system's fragment. As a first step, two reference systems are 
chosen such that the fragment center of mass is in the same position. This 
operation is equivalent to finding the translation between the template and the 
system. We then have two lists of atomic positions, $\{ \mathbf R^{T,S}_a\}$, 
where $T$ and $S$ label the template and system fragment respectively, and the 
subscript $a$, indicating the atom, ranges from 1 to $N$, the number  of atoms in 
the fragment.

If the system fragment is a \emph{rigid} displacement of the template (i.e.\ its 
internal coordinates are unchanged), the rotation matrix $\mathcal R$ we seek is 
such that $ \mathbf R^S_a = \sum_{b=1}^N {\mathcal R}_{ab} \mathbf R^T_b$.
In general, we should assume there is a slight modification of the internal 
coordinates, as the geometry of the fragment might be affected by the 
interaction with the environment.
In this case, the matrix $\mathcal R$ is such as to \emph{minimize} the cost 
function
\begin{equation}\label{wahbacost}
J(\mathcal R) = \frac{1}{2} \sum_{a=1}^N || \mathbf{R}^S_a - \sum_{b=1}^N {\mathcal 
R}_{ab} \mathbf R^T_b ||^2\;.
\end{equation}
The determination of the matrix $\mathcal R$ in such a manner constitutes a 
version of the well-known Kabsch algorithm~\cite{Kabsch} (also known as Wahba's problem)~\cite{Wahba2}, which can be solved by 
the Singular-Value Decomposition of the 3-by-3 matrix \cite{Wahba1} $\mathcal 
B_{ij} = \sum_{a=1}^N (R^S_a)_i (R^T_a)_j$. After having found two matrices 
$\mathcal U$ and $\mathcal V$ and a diagonal matrix $\mathcal S$  such that 
$\mathcal B = \mathcal U \mathcal S \mathcal V^t$, the optimal rotation is
\begin{equation}
 \mathcal R = \mathcal U \mathcal D \mathcal V^t\, \qquad \mathcal D = 
\mathrm{diag}\left(1,1,\det(\mathcal U) \det(\mathcal V) \right)\;.
\label{eq:R}
\end{equation}
The value of $J(\mathcal R)$ defined in \eqref{wahbacost} might then be used to 
quantify the validity of the rigid transformation approach. In the case where 
its value is below a given threshold (fixed to $10^{-3}$ in our case), we may 
proceed with the reformatting of the template basis functions, which will be 
denoted by $\Ket{\phi_\alpha^T}$ in what follows. 

%check goedeckerneelovMF is the correct reference
As described in Refs.~\cite{linear_paper, genovese:014109, goedeckerneelovMF}, 
from the expression of $\Ket{\phi_\alpha^T}$ in a Daubechies wavelets basis set, 
the so-called ``magic-filter'' transformation can be used to define a real space 
representation of the basis functions, given in terms of one-dimensional 
interpolating scaling functions (ISF)
\begin{equation}
 \phi_\alpha^T(x,y,z) = \sum_{i,j,k} c_{ijk} \; \varphi_i(x) \varphi_j(y) 
\varphi_k(z)\;,
\end{equation}
where $\varphi_i(x) \equiv \varphi(x/h - i)$ is one element of the ISF basis 
set, which is constituted of uniform translations of the mother function 
$\varphi(t)$ over the points of a uniform grid of spacing $h$, covering the 
entire simulation domain. %\luigi{should we put (again) a figure of ISF?}
 These points are labeled by indices $ijk$. The points $\mathbf r_{ijk} = ( h i, 
h j, h k)$ therefore lie within the box containing the support of 
$\phi_\alpha^T(\mathbf r)$.

This real-space expression is optimal in the sense that it preserves the same 
moments of the original representation given in Daubechies wavelets.
The interpolating property of the ISF basis set is such that $c_{ijk} = 
\phi_\alpha^T(\mathbf r_{ijk})$. 
Let us suppose we have a one-dimensional function expressed in ISF, namely $f(x) 
= \sum_i f_i \varphi_i(x)$.
We know that $f_i = f(h i)$. If we want to translate the function $f$ by a 
displacement $\Delta$, and express this function in the ISF basis, we have
$ f(x+\Delta) = \sum_i f'_i \varphi_i(x)$,
with 
\begin{equation}
f'_i = f(h i + \Delta) = \sum_j f_{i-j} t_{j}^\Delta\;,
\end{equation}
where the filter $t_{j}^\Delta = \varphi(j + \Delta/h)$ implements the (uniform) 
translation. 
This filter has a limited extension (the same as the function $\varphi(x)$) and of course 
$t_{j}^{h k} = \delta_{j,-k}$.

Imagine now we have a \emph{different} ISF basis set $\{\varphi_I(\tilde x) \}$ 
defined on a uniform grid spacing of 
separation $\tilde h$ and a reference frame $\tilde x_I= \tilde h I$, which is 
related to $x$ by a  more complicated transformation $\tilde x(x)$ of the 
coordinate space.
If this transformation can be inverted, by $x(\tilde x)$, then 
a new function $\tilde f(\tilde x) \equiv f(x(\tilde x))$ can be defined in this 
frame.
For each grid point $I$, it is then always possible to find $\bar i$ in the old 
frame such as to minimize the absolute value of $\Delta_I \equiv x(\tilde x_I) - 
h \bar i$.

Using the above relations we might approximate $\tilde f(\tilde x) \simeq \sum_I 
\tilde f_I \varphi_I(\tilde x)$, where
\begin{equation} \label{smoothtranslation}
\tilde f_I =\tilde f(\tilde x_I) = f ( h \bar i + \Delta_I) = \sum_j f_{\bar 
i-j} t_{j}^{\Delta_I}\;.
\end{equation}
If the transformation $\tilde x$ is a continuous function of $x$ which varies 
slowly enough, this is in general a rather good approximation (see Fig.~\ref{fig:rotating}).

This framework can be easily generalized to a roto-translation in three 
dimensions.
Indeed, we would like to estimate the function
\begin{equation}
 \phi_\alpha^S(\tilde {\mathbf r}) \equiv \phi_\alpha^T(\mathbf r 
(\tilde{\mathbf r})) \simeq \sum_{I,J,K} \tilde c_{IJK} \; \varphi_I(\tilde x) 
\varphi_J(\tilde y) \varphi_K(\tilde z)\;,
\end{equation}
where the coordinates $\tilde{\mathbf r} = ( \tilde x, \tilde y, \tilde z )$ are 
defined as
\begin{equation}
\tilde {\mathbf r} = \mathcal{R} \cdot \mathbf r \,
\end{equation}
where $\mathcal{R}$ is calculated by Eq.~\eqref{eq:R}.
In addition a rigid shift vector $\mathbf s = (s_x, s_y, s_z)$ is defined as the 
difference between the coordinates of the center of mass of the two fragments. 
If the rotation is the identity matrix, the template reference frame is then 
$\tilde {\mathbf r}=\mathbf r + \mathbf s$.
As in the one dimensional case presented above, the interpolation depends on the \emph{inverse} mapping~$\mathbf r (\tilde{\mathbf r})$. We detail in the following a procedure to identify such a function.

The coefficients $\tilde c_{IJK}$ of $\phi_\alpha^S(\tilde{\mathbf r})$ can be 
found in three steps. We first start by considering the transformation law for 
$\tilde x$. This transformation can be thought of as a function of the template 
coordinates $\mathbf r$:
\begin{equation}\label{xtilde}
\tilde x(x,y,z) = \mathcal R_{11} x  + \mathcal R_{12} y + \mathcal R_{13} z \;.
\end{equation}
In the same spirit as Eq.~\eqref{smoothtranslation}, we may invert 
Eq.\eqref{xtilde} with respect to one template coordinate $t = x ,y ,z$ into 
$\tilde x$ in the system's reference frame. 
The choice of the variable $t$ depends on the entries of the rotation matrix, and it 
is in general given by the coordinate which is multiplied by the coefficient of the
highest absolute value in Eq.~\eqref{xtilde}. 
This choice guarantees that the $t$ variable is the one for which $\tilde x - t$ 
is slowly varying.
Let us imagine $t=x$ for this example. We can define the function
\begin{multline}
 \phi_\alpha^{(1)}(\tilde x, y, z) = \phi_\alpha^T(x(\tilde x, y, z)-s_x, y, z) 
\\ = \sum_{I,j,k} \tilde c_{I,j,k}  
 \varphi_I(\tilde x) \varphi_j(y) \varphi_k(z)\;,
\end{multline}
by proceeding for all $j,k$, as described in Eq.\eqref{smoothtranslation}, to 
define the coefficients $\tilde c_{I,j,k}$.
The second step is related  to the expression of $\tilde y$.
Depending on the choice of the variable $t$ in the first step, we have to 
consider one of these three relations:
\begin{align}
\mathcal R_{11} \tilde y & =\mathcal R_{21} \tilde x+ \mathcal R_{33} y - 
\mathcal R_{32} z \label{ytilde}  \;, \\
\mathcal R_{12} \tilde y &= \mathcal R_{22} \tilde x - \mathcal R_{33} x + 
\mathcal R_{31} z  \;, \\
\mathcal R_{13} \tilde y &= \mathcal R_{23} \tilde x + \mathcal R_{32} x - 
\mathcal R_{31} y \;,
\end{align}
which hold when in the first step $t=x,y,z$ respectively.
These relations can be derived using the orthogonality of the rotation matrix~$\mathcal{R}$.
This function can now be inverted with respect to one of the old variables.
Again, this choice will depend on the values of the coefficients multiplying each 
variable. 

In our example, we have to consider the relation~\eqref{ytilde} as we have chosen $t=x$ in the first step.
We choose to invert the relation with respect to $z$, having 
$z=z(\tilde x,\tilde y, y)$.
In this case we will have, as a second step
\begin{multline}
  \phi_\alpha^{(2)}(\tilde x, \tilde y, y) =  \phi_\alpha^{(1)}(\tilde x, y, 
z(\tilde x, \tilde y, y)-s_z) \\=
  \sum_{I,J,j} \tilde c_{I,j,J}  
 \varphi_I(\tilde x) \varphi_J(\tilde y) \varphi_j(y)\;.
\end{multline}
In the third step, the remaining variable, (which is $y$ for the illustrated 
example), can be directly obtained from the inverse relation 
\begin{equation}
 \mathbf r = \mathcal R^{-1} \cdot \tilde{\mathbf r} = \mathcal R^t \cdot 
\tilde{\mathbf r}\;,
\end{equation}
which is easier to express as $\mathcal R$ is an orthogonal matrix.
In our case, the final result is therefore
\begin{equation}
  \phi_\alpha^{S}(\tilde x, \tilde y, \tilde z) =   \phi_\alpha^{(2)}(\tilde x, 
y(\tilde x, \tilde y, \tilde z)-s_y, \tilde z) \;.  
\end{equation}
We recall that the definition of $\phi_\alpha^{(1,2)}$ depends on the order of 
the operation. Here we have chosen to interpolate first with respect to $x$, then $z$ and $y$.
The best choice of order depends only on the entries of the matrix $\mathcal R$.

\subsubsection{Accuracy}

In order to assess the accuracy of the reformatting scheme, we have applied it 
to a water molecule undergoing a series of rotations.  Support functions were 
generated for a template water molecule, using a dense grid with a spacing of 
$0.132~$\AA, and were then reused for water molecules in a variety of 
different orientations using a less dense grid with a spacing of $0.185$~\AA.  
As a point of comparison, calculations were also performed for each orientation 
fully optimizing the support functions with a grid spacing of $0.185$~\AA. 
This allows us to quantify
both the error introduced by the support function reformatting and the errors 
due to representing the wavefunctions on a fixed grid, i.e.\ the so-called 
`eggbox effect'.  The eggbox effect of the standard cubic scaling 
approach is also presented. The computational setup has been chosen such that the difference in 
ground state energies between the cubic and support function approaches is of the order of 1~meV/atom.

\begin{figure}
\includegraphics[width=0.45\textwidth]{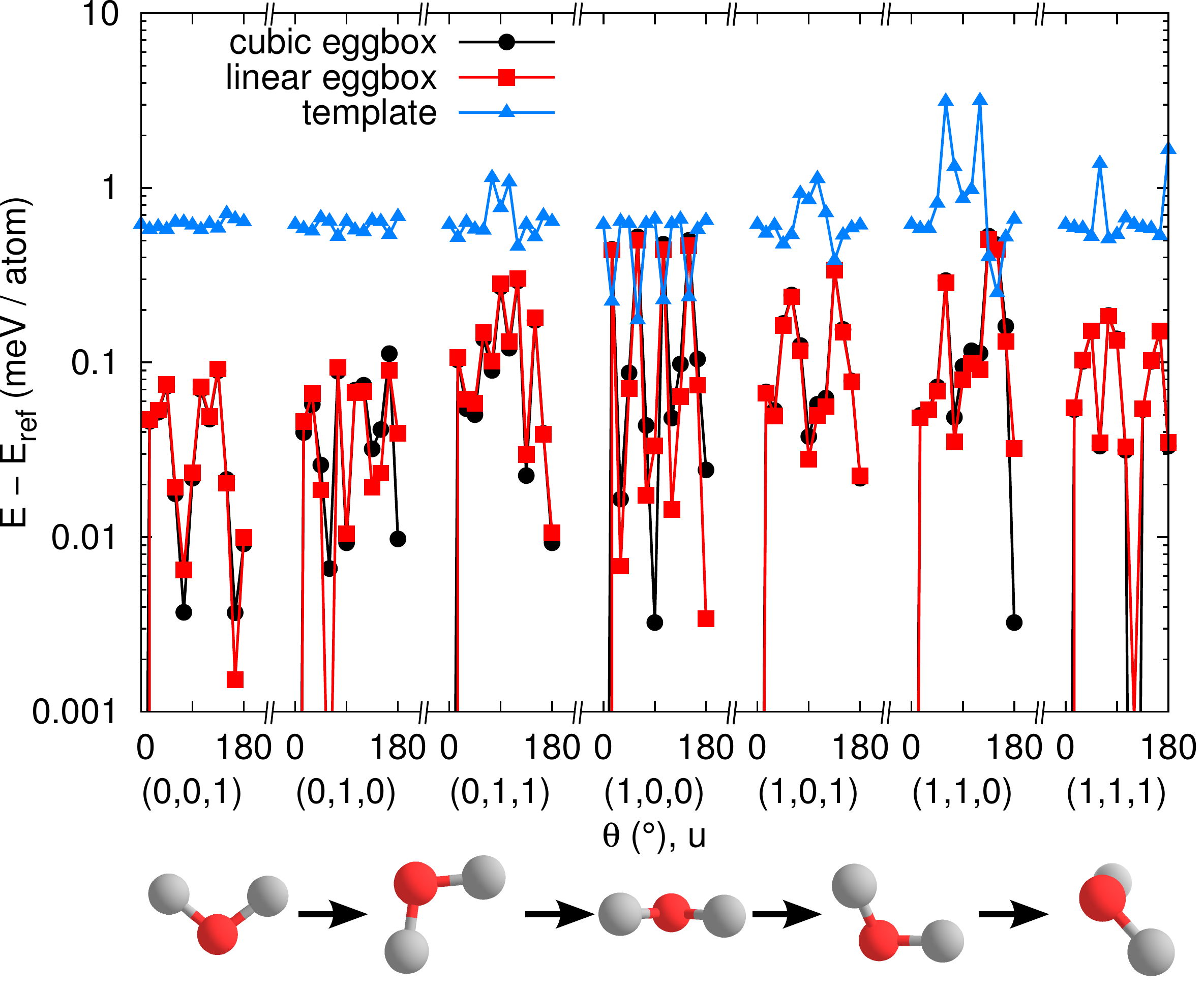}
\caption{Plot showing the energy variation for a water molecule rotated 
through different angles ($\theta$) and axes of rotation ($u$).
Results are shown for the standard cubic scaling approach (`cubic eggbox'),
fully optimized support functions (`linear eggbox') 
and a fixed support function basis generated for a template molecule (`template'). 
The cubic reference is the energy at the initial orientation calculated using the cubic approach,
for the linear and template approaches
it is the same quantity calculated in the fully optimized support function basis.
There is a roughly constant error of 1~meV/atom in the support function basis compared to the cubic
scaling approach.
Selected orientations are shown along the bottom.}\label{fig:rotating}
\end{figure}

The results are shown in Fig.~\ref{fig:rotating}, where we can see that the 
eggbox effect is of the order of 0.1~meV/atom. As both the cubic and linear scaling approaches
use the same underlying grid, the variation is similar in each case.
The error due to the interpolation also remains small -- less than a few 
meV/atom.  Importantly, the overall error for the reformatted calculations remains of the same order of magnitude 
as that due to the selected localization radii of the support functions. 

\section{Results}

Below we present results for three different systems, where the first two can be 
validated against CDFT implementations in other codes.  In each case we use the 
local density approximation (LDA) exchange-correlation 
functional~\cite{PhysRevLett.45.566} and HGH 
pseudopotentials~\cite{PhysRevB.58.3641} within isolated boundary conditions.
For carbon, nitrogen and oxygen we use four support functions per atom,
for hydrogen we use one, and for zinc we use nine.

\subsection{$\textrm{N}_2$}

\begin{figure}
\includegraphics[scale=0.35]{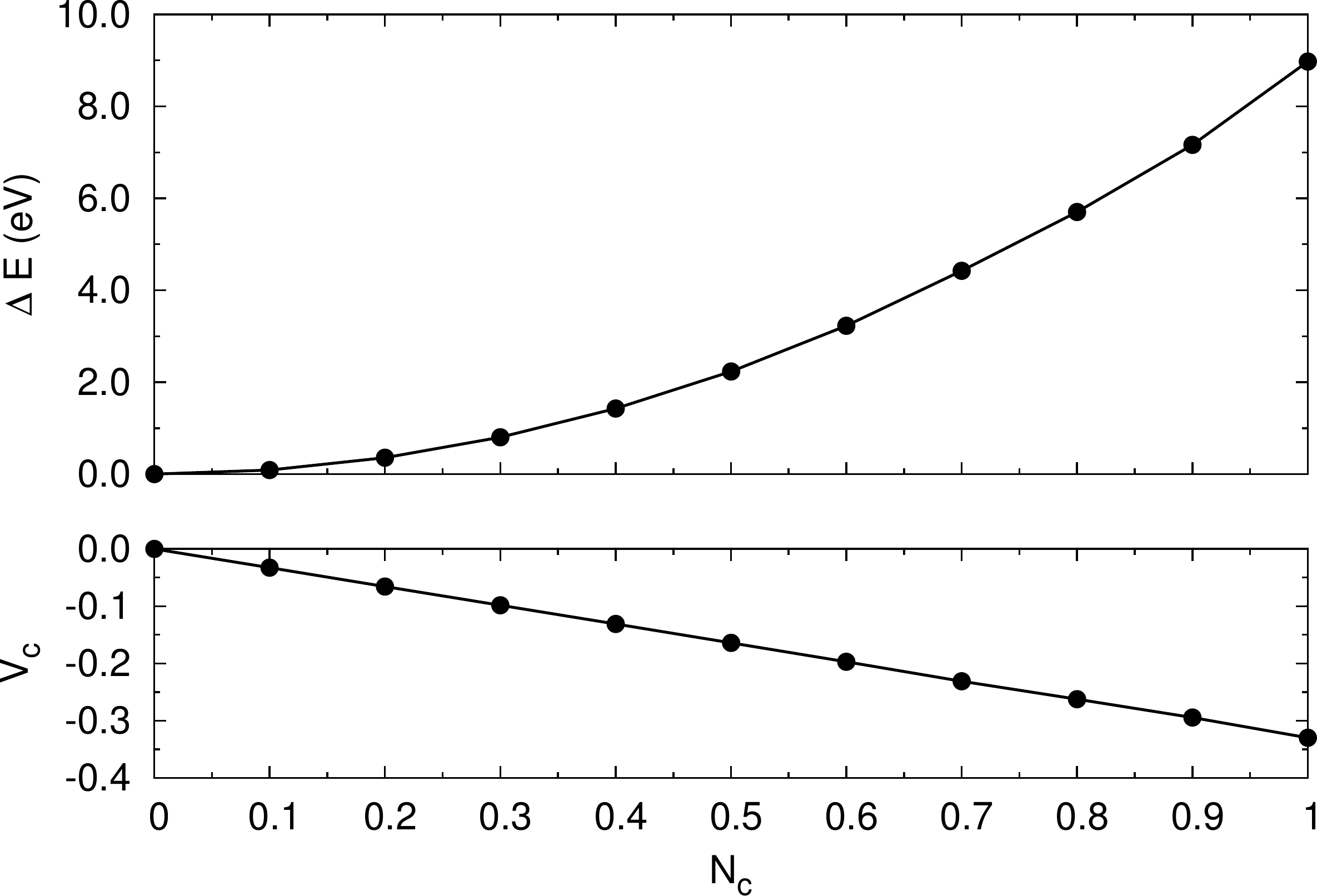}
\caption{Change in energy with respect to an unconstrained calculation and 
applied potential value for differing charge separations in 
$\textrm{N}_2$.}\label{fig:n2}
\end{figure}

Wu and Van Voorhis have previously studied 
$\textrm{N}_2$~\cite{doi:10.1021/ct0503163} and so this system provides a useful 
test case.  We used a grid spacing of $0.185$~\AA\ with support function
radii of $7.4$~\AA, i.e.\ completely filling the simulation cell; here 
we aim to validate only the general correctness of the implementation of CDFT 
rather than the full fragment approach.  Fixing the bond length at 1.12~\AA, we 
have varied the charge separation between the two atoms, results for which are 
shown in Fig.~\ref{fig:n2}.  For simplicity, the calculations were performed 
only in the spin-averaged sense.  Our results are closer to those obtained by Wu 
and Van Voorhis using a Becke weight population than the L\"{o}wdin scheme, 
however given that we have used the LDA whereas they used B3LYP we do not expect 
exact agreement.  Furthermore, we should recall that the fragment approach presented here is aimed at 
systems where the donor and acceptor are well separated, whereas the use of 
support functions optimized for an isolated nitrogen atom is necessarily an 
approximation in this case.  Nonetheless, we have successfully reproduced the correct 
trends for both the energy and the Lagrange multiplier.

\subsection{ZnBC}

\begin{figure}
\includegraphics[width=0.45\textwidth]{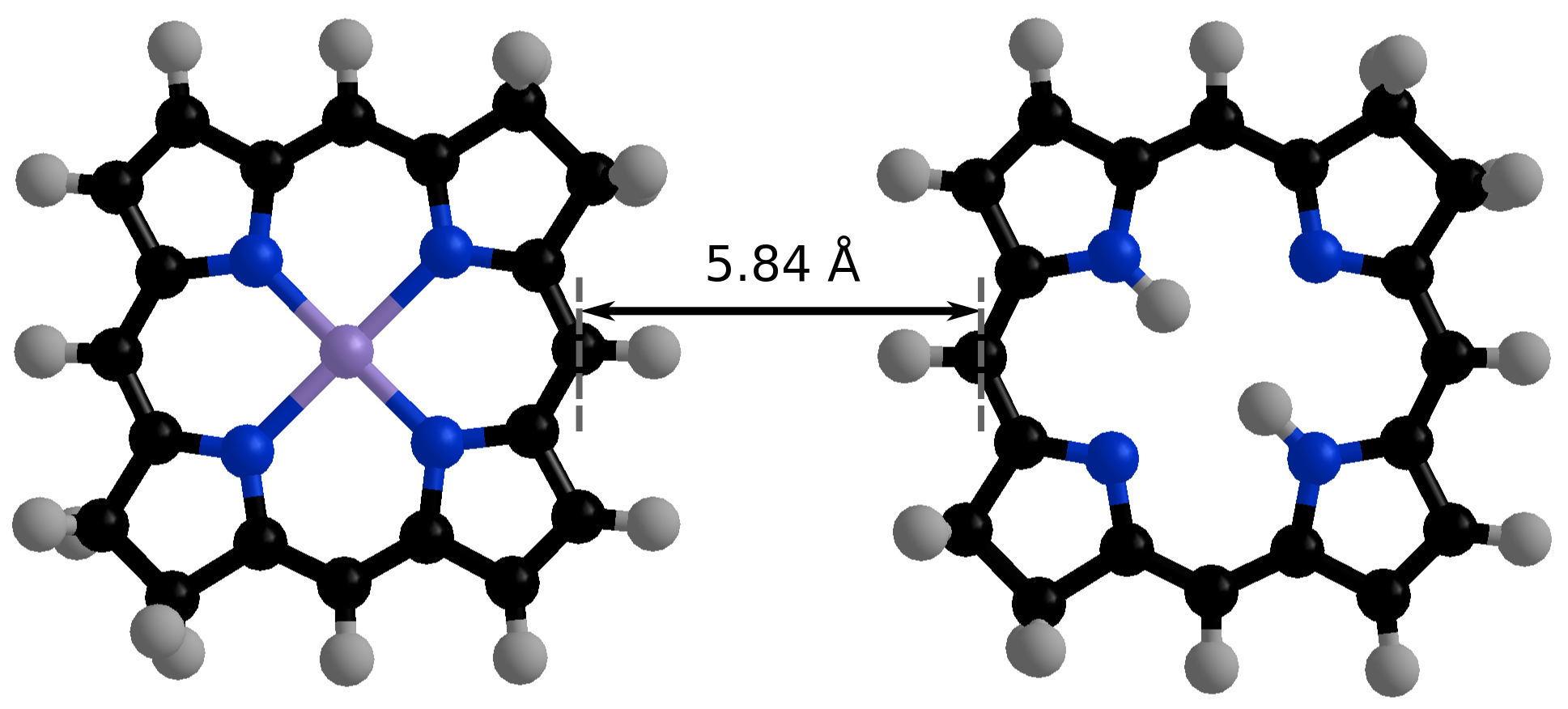}
\caption{The ZnBC-BC model complex.}\label{fig:znbcl}
\end{figure}

As a more elaborate test case we take the zincbacteriochlorin-bacteriochlorin 
(ZnBC-BC) complex, which has also been studied previously in some detail, both 
with CDFT~\cite{doi:10.1021/ct0503163,PhysRevA.72.024502} and other 
approaches, e.g.\ 
Refs.~\onlinecite{PhysRevLett.109.167801,doi:10.1021/ja039556n}. 
This system is ideally suited to 
our approach as the donor and acceptor are clearly separated.  
Furthermore, TDDFT has been shown to give incorrect energies for the 
$\textrm{ZnBC}^+\textrm{-}\textrm{BC}^-$ and 
$\textrm{ZnBC}^-\textrm{-}\textrm{BC}^+$ charge transfer (CT) excited 
states~\cite{doi:10.1021/ja039556n} and so the advantages of CDFT are clear.
It has previously been demonstrated that the differences between a 
(1,4)-phenylene-linked ZnBC-BC complex and a model complex where the link is 
eliminated are small~\cite{doi:10.1021/ja039556n}; for simplicity we 
therefore choose to use the latter, where the distance between the two previously linked 
carbon atoms is 5.84\AA, as depicted in Fig.~\ref{fig:znbcl}.   
Taking the coordinates from Ref.~\onlinecite{doi:10.1021/ja039556n}, we relaxed 
the isolated ZnBC and BC molecules separately, then built
the model complex without further relaxation.
We used a grid spacing of $0.185$~\AA\ and localization radii of 
$5.82$~\AA.  
To assess the accuracy of the fragment support functions we compare the neutral energies
for the model complex with those obtained using cubic scaling BigDFT.
The results are shown in 
Tab.~\ref{table:znbc}, where we can see that the error for both the model
complex and the isolated molecules is less than 1~meV/atom.

\begin{figure}
\centering
\subfigure[$\textrm{ZnBC}^-\textrm{-}\textrm{BC}^+$]{\includegraphics[width=0.45\textwidth]{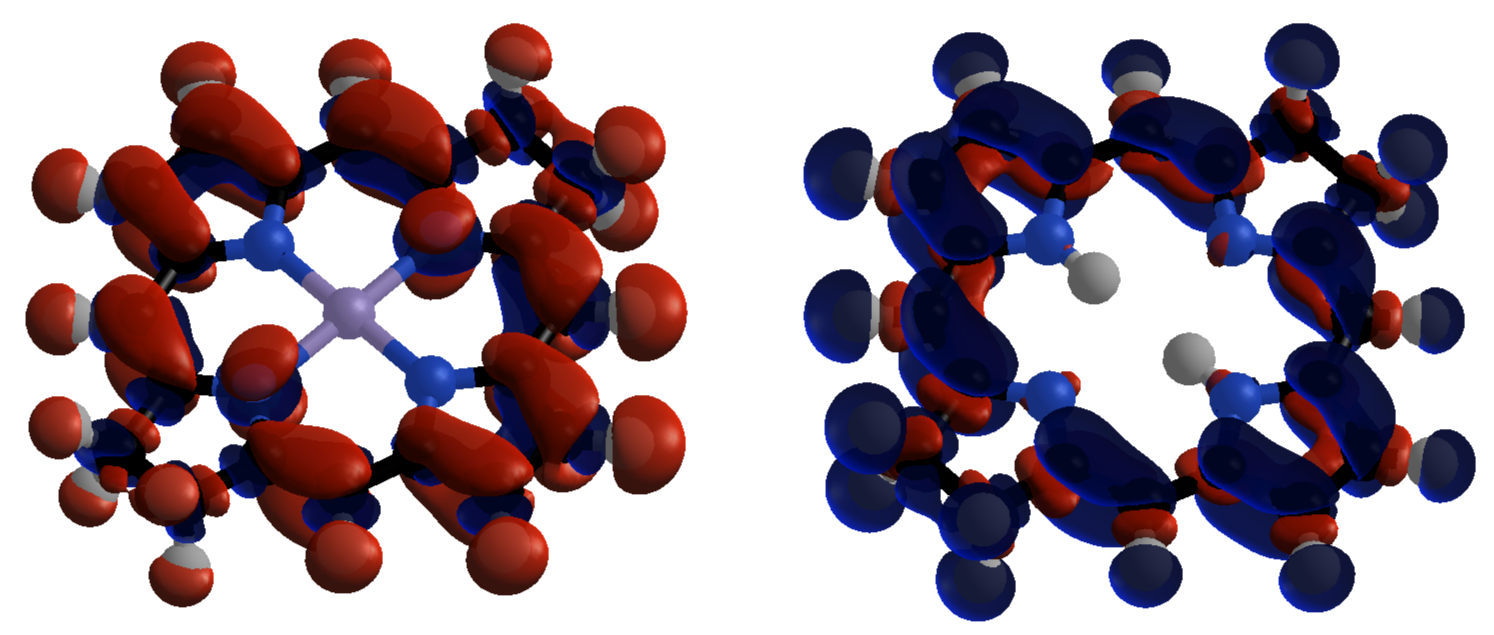}}
\subfigure[$\textrm{ZnBC}^+\textrm{-}\textrm{BC}^-$]{\includegraphics[width=0.45\textwidth]{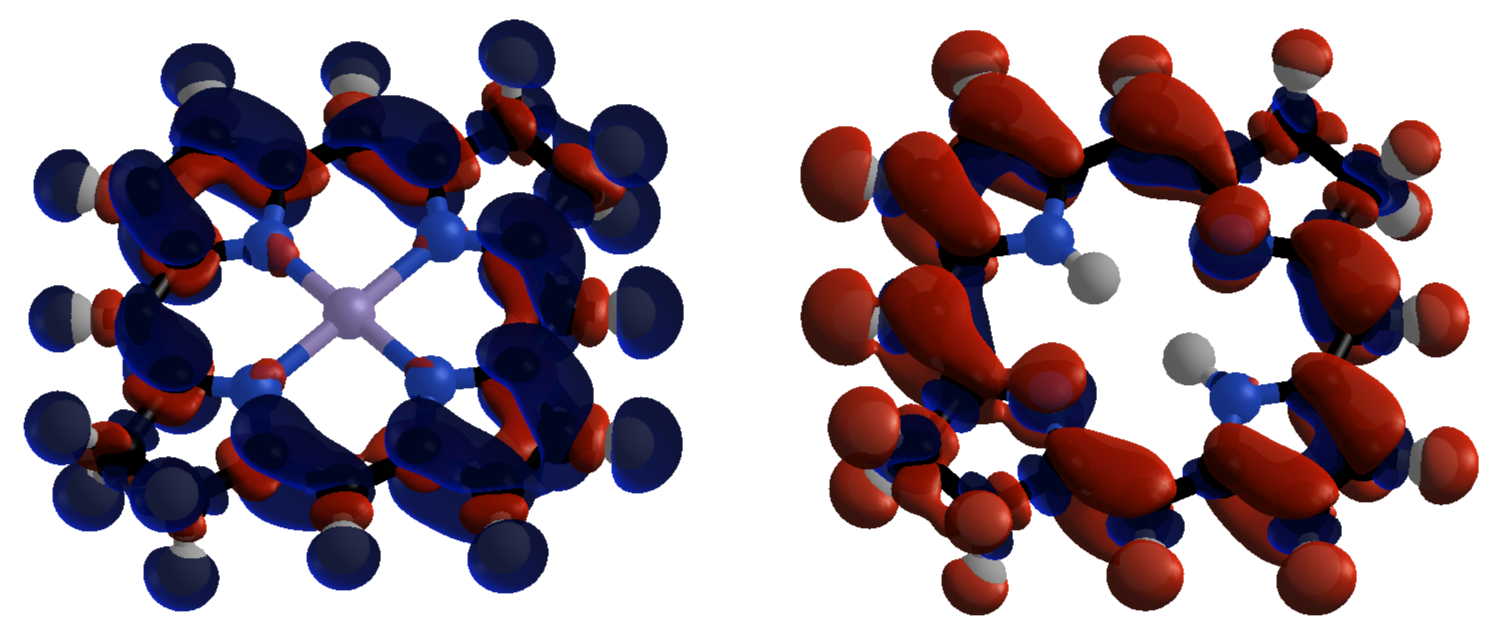}}
\caption{Density differences between the neutral and charged calculations for the two charge transfer states.  Red (blue) indicates an increase (decrease) in the electronic charge density with respect to the neutral.}\label{fig:densplots_znbcl}
\end{figure}

The energies for the two CT excited
states relative to the unconstrained DFT ground state are 3.71~eV for 
$\textrm{ZnBC}^+\textrm{-}\textrm{BC}^-$ and 3.98~eV for 
$\textrm{ZnBC}^-\textrm{-}\textrm{BC}^+$, which is consistent with previous 
results~\cite{doi:10.1021/ct0503163,PhysRevA.72.024502}. 
We can also gain some insight into the nature of these CT states by plotting
the difference in the electronic density between the neutral and constrained calculations,
as in Fig.~\ref{fig:densplots_znbcl}.  Not only is the charge transfer characteristic clear,
the plot for $\textrm{ZnBC}^+\textrm{-}\textrm{BC}^-$ also shows remarkably good agreement
with previous calculations that used the significantly more expensive 
Bethe-Salpeter approach~\cite{PhysRevLett.109.167801}, which 
confirms that CDFT can be used to obtain physically relevant CT excitons, and provide a reliable estimation of the corresponding excitation energies.

\begin{table}
\centering
\begin{tabular}{lccc}
\hline\hline\\[-3.0ex]
& cubic & frag. & diff.\\
\multirow{2}{*}{ } & \multicolumn{2}{c}{ (eV)} & (meV)\\
\hline\\[-1.0ex]
ZnBC & $-4472.626$ &  $-4472.604$ & $22.6$ \\[0.5ex] %22.581
BC & $-4471.998$ &  $-4471.980$ & $17.7$\\[0.5ex] % 17.656
ZnBC-BC & $-8944.629$ & $-8944.575$ & $54.1$ \\[0.5ex] %54.071
$\textrm{ZnBC}^-\textrm{-}\textrm{BC}^+$ & - & $-8940.592$  & - \\[0.5ex]
$\textrm{ZnBC}^+\textrm{-}\textrm{BC}^-$ & - & $-8940.860$ & - \\[0.5ex]
\hline\hline
\end{tabular}
\caption{Energies for isolated ZnBC and BC, the neutral model ZnBC-BC complex 
and the two lowest energy CT states, as calculated using standard BigDFT 
(`cubic') and the fragment approach (`frag.').  Where applicable the 
difference between the two approaches is also indicated (`diff.').}  
\label{table:znbc}
\end{table}

\begin{figure}
\includegraphics[scale=0.35]{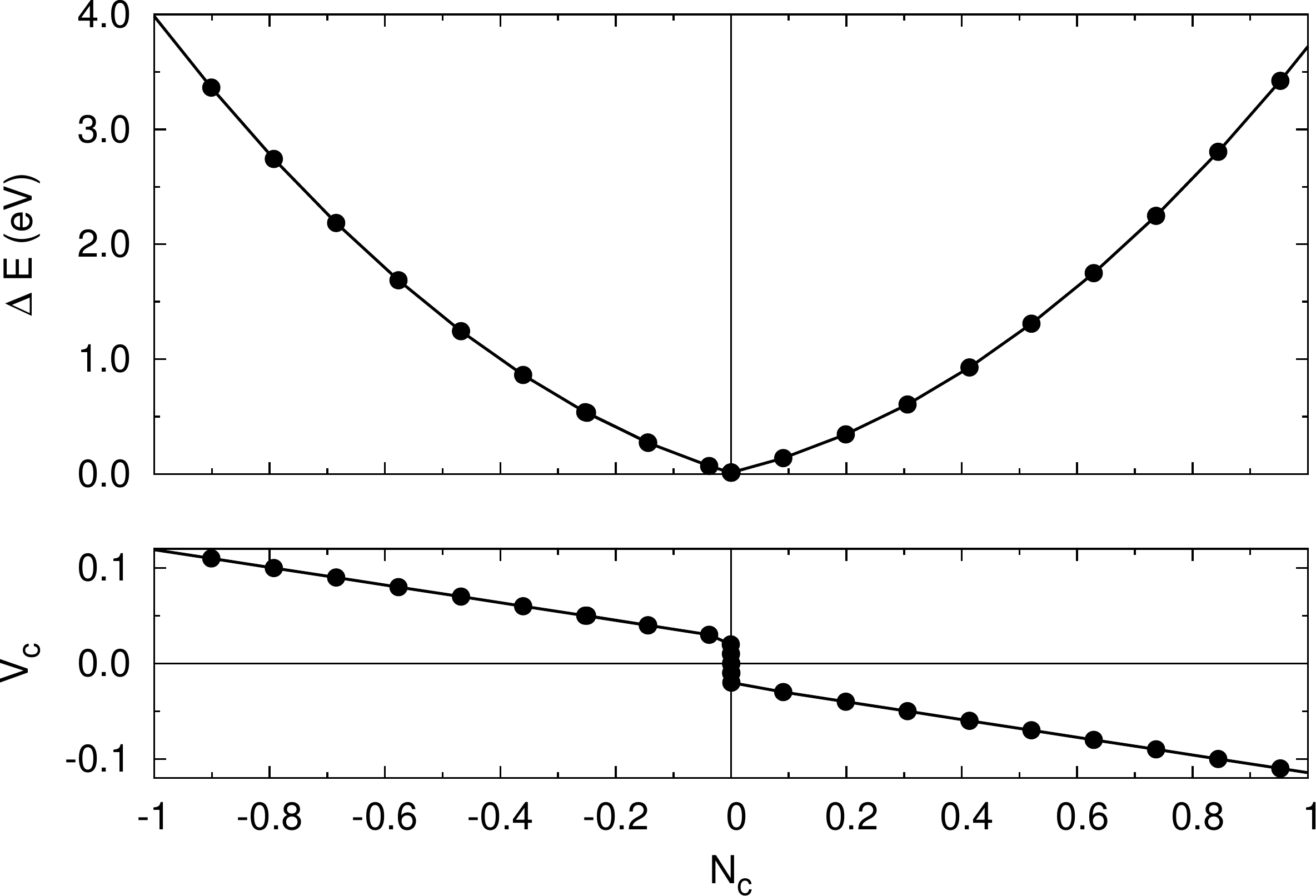}
\caption{Applied potential value and change in energy compared to an 
unconstrained calculation for differing charge separations in the ZnBC-BC model 
complex.}\label{fig:znbcl_plot}
\end{figure}

We have also plotted the relationship between the constraining potential, $V_c$,
the total energy relative to the unconstrained calculation, 
$\Delta E$, and the charge difference between the two molecules, $N_c$.
This is shown in Fig.~\ref{fig:znbcl_plot}, where 
$N_c=1$ corresponds to $\textrm{ZnBC}^+\textrm{-}\textrm{BC}^-$ and $N_c=-1$ 
corresponds to $\textrm{ZnBC}^-\textrm{-}\textrm{BC}^+$; our results agree well 
with previous calculations~\cite{doi:10.1021/ct0503163,PhysRevA.72.024502}, 
despite the use of a different exchange-correlation functional.
This test highlights the robustness of the method -- in order for the correct 
value of $V_c$ to be found within a minimal number of iterations of the 
constraint loop, there should be a smooth relationship between a given
$V_c$ and the resulting $N_c$.  If for certain values of $V_c$ the convergence 
is insufficient, such that the final charge deviates from the correct 
value, this will negatively impact the search for the correct $V_c$.  We 
observed that in general such a smooth curve is straightforward to obtain, 
given a reasonable initial guess for the density kernel and therefore charge 
density.  As discussed in Section~\ref{sec:cdft}, this can be achieved by defining
the initial occupancies in a manner which is consistent with the desired charge difference.

\subsection{$\textrm{C}_{60}$}
In order to accurately calculate material properties it is important to account for environmental effects, 
e.g.\ by including a solvent or neighboring molecules in a molecular material.  
However, this can considerably increase the cost of a simulation, 
as in the case of large systems in solution where the solvent must fill a correspondingly large volume.  
Various strategies have been developed for reducing the cost, for example by 
using implicit solvation 
methods~\cite{doi:10.1021/cr00031a013,jcp/124/7/10.1063/1.2168456,
FossoTande2013179,0295-5075-95-4-43001}, however it is frequently desirable to 
treat explicitly the environmental degrees of freedom.  Thanks to the fragment approach, the treatment of 
solvents and other surrounding molecules can readily be achieved in BigDFT with relatively low cost, as we will 
demonstrate through the example of the fullerene $\textrm{C}_{60}$ in two different environments:
when in an aqueous solution and when surrounded by other $\textrm{C}_{60}$ molecules.
For each system we constrain a charge of $\pm 1$ to 
the central $\textrm{C}_{60}$ molecule in order to determine the environmental 
impact on the ionization potential (IP)
and electron affinity (EA).

For traditional DFT calculations with semilocal functionals like LDA it is well known that the above quantities are badly estimated by the frontier orbitals, i.e.\ the HOMO (LUMO) for the IP (EA).
%(optical) band gap is underestimated by the LDA:
%in this case the isolated value calculated using the cubic scaling approach is 1.6~eV compared to 
%the experimental value of 2.5~eV~\cite{PhysRevLett.68.1232}.
Therefore, in order to extract physically meaningful information, one must either use more expensive beyond-DFT
approaches, or instead calculate the IP and EA using the so-called $\Delta$SCF method.  
This is made possible when explicitly charged calculations are available, i.e.\ when charged and neutral calculations have energies that can be measured with respect to a common reference. The treatment of the electrostatic potential which is included in the BigDFT code makes such a comparison possible \cite{PSWires}.
This latter approach results in values which match experiment much better than traditional semilocal functionals,
indeed our results for the IP and EA
of the isolated molecule agree very well with the experimental values of 
7.6~eV~\cite{A1991EV68500010,A1992HB67000001,459778} and 2.7~eV~\cite{478732} respectively.
In this case, we wish to apply the $\Delta$SCF approach to a molecule in an environment, which, as we will show,
is easily achieved using CDFT. 

On the other hand, with unconstrained DFT calculations the use of the $\Delta$SCF approach is much more delicate when studying environmental effects with LDA: as the charge tends to be overly delocalized, charged calculations do not simply represent a perturbation from the isolated values, as discussed in more detail below.   
%with cas the charged system on the other hand, 
%it is clear from the large difference with the isolated values
%that, unlike the constrained calculations, the results for the shorter distances 
In other words, the calculated energy differences do not correspond to the IP and EA
of $\textrm{C}_{60}$ in an environment, but to a completely different quantity.
If one wishes to calculate this quantity it is therefore essential to use CDFT.

\subsubsection{Computational details}
\begin{figure*}
\centering
\subfigure[The three configurations A, B and C (left to right) of 
$\textrm{C}_{60}$ in $\textrm{H}_{2}\textrm{O}$.]{\includegraphics[width=0.7\textwidth]{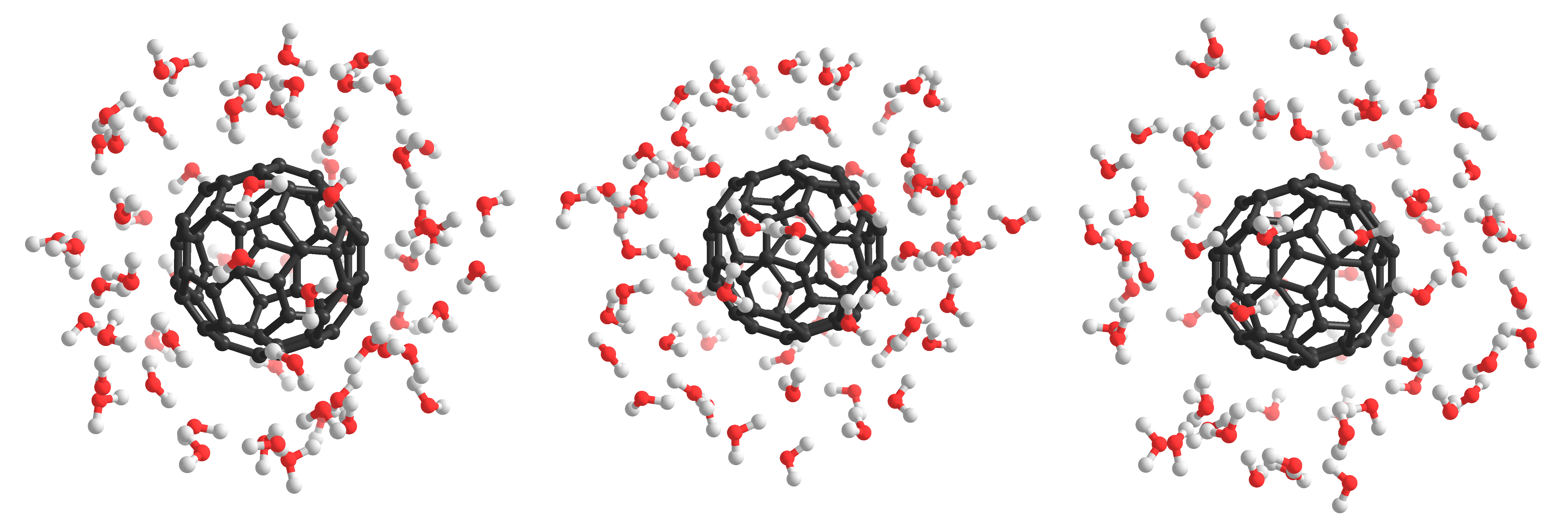}\label{fig:c60h2o}}
\subfigure[$\textrm{C}_{60}$ with its six nearest neighbors. The blue lines are drawn between the molecular centers along the axes.]{\includegraphics[scale=0.25]{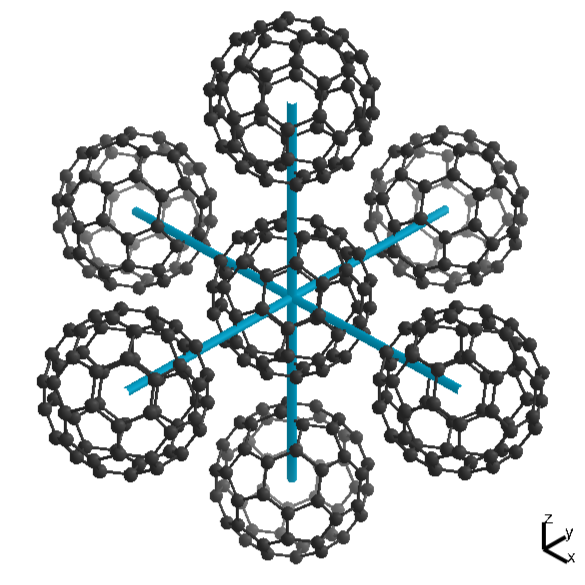}\label{fig:c60_struc}}
\caption{The different environments for $\textrm{C}_{60}$.}
\end{figure*}

There have been a number of previous studies of $\textrm{C}_{60}$ in water, both 
experimental and 
theoretical~\cite{doi:10.1021/nl070308p,Rivelino20062925,Scharff20041203,
jcp/123/20/10.1063/1.2121647,C3CP50187F,jcp/138/4/10.1063/1.4789304,
jcp/125/3/10.1063/1.2217442}, however they have mainly focused on neutral 
fullerenes.
Previous research has indicated the existence of a first hydration shell surrounding 
$\textrm{C}_{60}$ containing between 60 and 65 water 
molecules~\cite{Rivelino20062925,doi:10.1021/nl070308p,Scharff20041203}, 
we have therefore chosen to restrict ourselves to systems containing 66 water molecules.  
We present results for three example structures, which are depicted in 
Fig.~\ref{fig:c60h2o}.  They were generated by inserting the $\textrm{C}_{60}$ into 
water droplets where the water molecules were deposited with random orientations 
at random positions subject to the room-temperature density of water.  The 
structures were then relaxed until the RMS forces were below 10~meV/\AA.
For the environment of fullerenes, we limit the cost of the simulations by including six nearest
neighbor fullerenes only, so that the system is arranged as a three dimensional cross, 
as depicted in Fig.~\ref{fig:c60_struc}.
Each of the fullerenes was considered in its gas-phase structure.

The fragment calculations were performed with a grid spacing of $0.185$~\AA, while the 
template calculations were performed using a denser grid of $0.132$~\AA\ to ensure 
accurate reformatting; we used support function radii of $4.23$~\AA. These values have been chosen such as to ensure the applicability
of the L\"{o}wdin approach for the weight matrix on the central \ce{C60} whilst preserving absolute accuracy of the unconstrained calculations to the order of \unit[3]{meV/atom}, see data in Sec.~\ref{testunconstrained}. 
%for the fullerene environment tests were also performed with radii of $3.18$~\AA\
%and $5.29$~\AA\ -- for the smaller radius the localization constraint had a 
%detrimental effect on the accuracy of the calculations, whereas the energy differences for the larger radius
%were very similar to those at $4.23$~\AA.  For very large radii 
%the accuracy of the L\"{o}wdin approach for calculating the weight matrix will 
%decline but these tests confirm we are well within the region where it is applicable.
In order to ensure the support 
functions are sufficiently accurate for the negatively charged calculations for $\textrm{C}_{60}$,
the template calculation was performed optimizing three additional states (to 
account for degeneracies).

We continue to use the LDA functional as the difference between the LDA and other treatments like PBE
for the IP and EA of fullerenes has previously 
been shown to be negligible, and reasonable agreement with experiment has also been 
observed~\cite{:/content/aip/journal/jcp/129/8/10.1063/1.2973627}.  
We also neglected the modelling of dispersive terms on the \ce{C60} -- \ce{C60} interactions due to their negligible impact on frontier orbital eigenvalues and on total energy differences in charged calculations.
%It should be noted that in order to correctly model the interactions between the neighboring fullerenes dispersion effects should also be included, either by using a different functional or by applying a post-processing correction scheme.  However, our aim is to demonstrate the applicability of our approach to problems of this type and to provide preliminary numerical results as a starting point for future more detailed investigations; for this purpose, the LDA is sufficient.

\subsubsection{Testing the fragment approach}\label{testunconstrained}

The results for the $\textrm{C}_{60}$ structure with a center to center distance
of 10~\AA\ are shown in Tab.~\ref{table:c60} with the corresponding values for the
isolated molecule.  We also include the cubic scaling results as they
allow us to assess the accuracy of the fragment
approach for this system.  As anticipated, for the isolated molecule the fragment 
error is of the order of 0.2~eV, which is about \unit[3]{meV/atom}. 
%for the full system the error in total energies will be 
%greater, however we should also benefit from error cancellations so that the
%errors in the IP and EA will remain similar.
In order to confirm that this accuracy is preserved, we also compared unconstrained cubic and fragment calculations for the seven $\textrm{C}_{60}$
structure.  %CDFT has not been implemented for the cubic approach so we can only compare unconstrained calculations.
%As discussed in Section~\ref{sec:frag}, the final solution found by the fragment approach can depend strongly on the initial guess, therefore it is important that the initial charge distribution is close to the required distribution, i.e.\ in this case the excess charge should be located on the central molecule.  
%In order t
To find an unconstrained solution we built the initial guess from the fragment densities in such a manner
that the charge was equally distributed among fragments; the final solution remained close to this charge distribution.
%This cannot be guaranteed to be the same
%solution as that found by the cubic approach where the %charge is in general less evenly distributed,
%nonetheless it should be similar enough to allow for a comparison.  Indeed, with the fragment approach 
We found values of
-3.681~eV and 6.707~eV, i.e.\ the difference with the cubic results is of the same magnitude as for the isolated molecule (see Tab.~\ref{table:c60}).

\begin{table}
\centering
\begin{tabular}{c|cc|ccc|cc}
\hline\hline%\\[-2.0ex]
\rule{0pt}{12pt} 
 & \multicolumn{2}{c|}{{isolated}} & \multicolumn{3}{c|}{in $\textrm{H}_{2}\textrm{O}$} &\multicolumn{2}{c}{in $\textrm{C}_{60}$ (10~\AA)}   \\ 
{$Q$}  & cubic & frag. &  A & B & C & cdft & cubic \\
\hline%\\[-2.0ex]
\rule{0pt}{15pt} 
$-1$ & $-2.795$ & $-2.589$ & $-2.017$ & $-2.728$ & $-2.180$ & $-2.854$ & $-3.803$  \\[0.5ex] 
$+1$ & $7.648$ & $7.783$ & $7.262$ & $8.033$ & $7.837$  & $7.526$ & $6.685$\\[0.5ex]
\hline\hline
\end{tabular}
\caption{Energy differences with neutral, i.e.\ $E^{Q}-E^0$ for $\textrm{C}_{60}$ when isolated and in 
the two environments.  Two values are given for the 
isolated $\textrm{C}_{60}$: that of the fragment approach, which in this case 
merely refers to a fixed support function basis as only one fragment is 
present, and the  cubic scaling reference.  
For the 
results in water, constrained fragment results are given. For the nearest neighbor results
(`in $\textrm{C}_{60}$'), results are presented for both the constrained fragment
and (unconstrained) cubic approaches. The unconstrained results exhibit stronger deviation from the isolated values, showing that the environment is not correctly modeled as it is not acting as a perturbation of the system. 
Units are in eV.}  
\label{table:c60}
\end{table}

\begin{figure}
\includegraphics[scale=0.35]{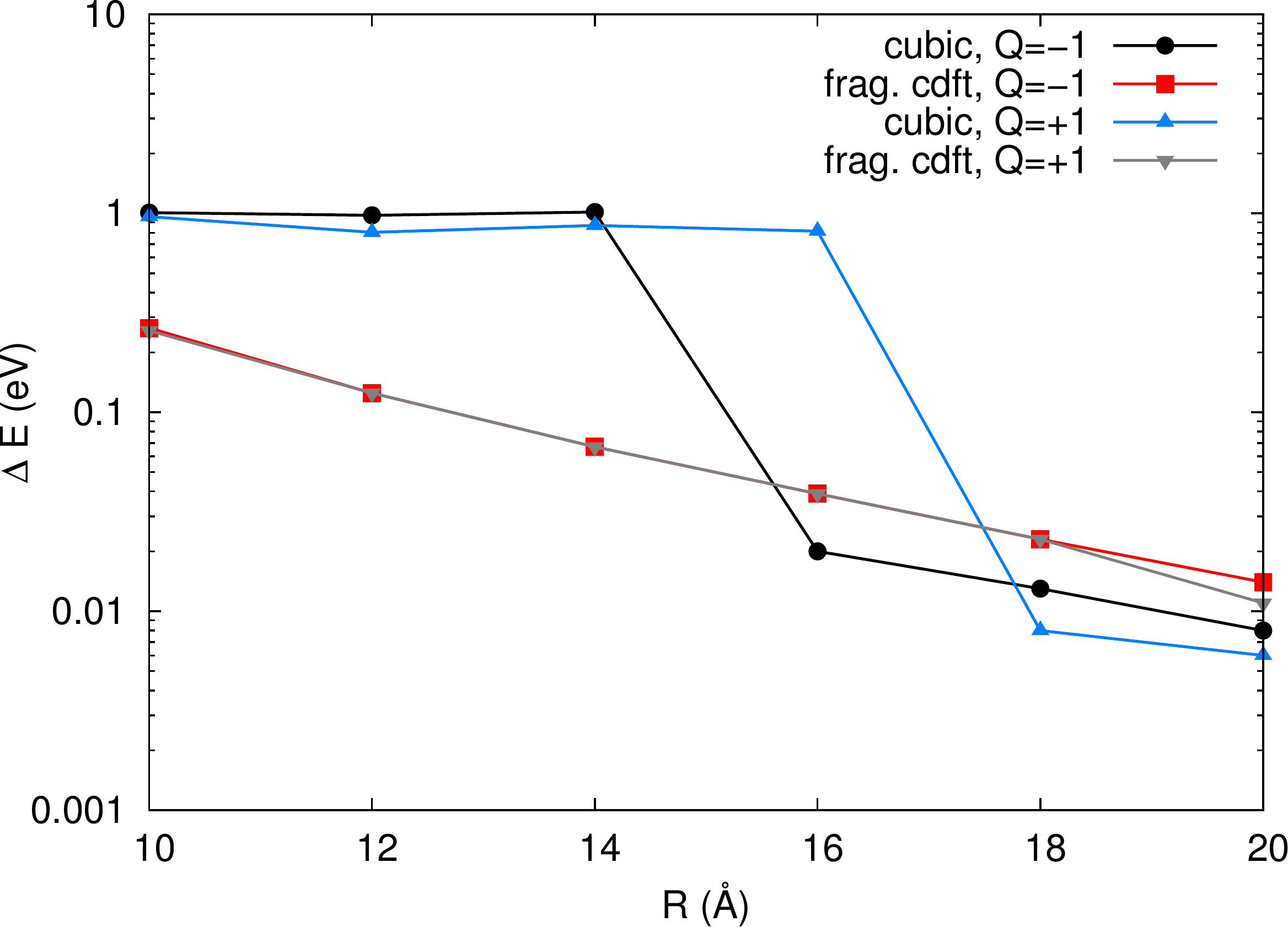}
\caption{Variation in electron affinity and ionization potential for increasing separations, where the distance is measured
between the centers of neighboring molecules.  The energy plotted is relative to the isolated value in the respective
basis, i.e.\ $\Delta E=(E^Q_{\textrm{isol.}}-E^0_{\textrm{isol.}})-(E^Q_{\textrm{full}}-E^0_{\textrm{full}})$,
where $Q$ indicates the charge state, `full' refers to the 7 molecule system and `isol.' refers to the isolated molecule.}\label{fig:c60_distance} 
\end{figure}

We have also investigated the effect of varying the separation between the molecules by
repeating the constrained fragment and (unconstrained) cubic calculations with center to center separations ranging from 10~\AA\ to 20~\AA, which corresponds to a shortest distance between molecules of 3.1~\AA\ to 13.1~\AA.  The results are plotted
in Fig.~\ref{fig:c60_distance}.  As expected, for large separations the results tend towards the isolated values.  
For the unconstrained calculations there is an abrupt change between two different states, whereas
with CDFT there is not only a smooth trend, but also an exponential relationship with distance, proving
that the fragment approach is sufficiently precise to capture such trends.  

Thus far, we have only considered calculations with a \emph{shifting} of the template support functions, however we also wish to demonstrate the effectiveness for \emph{rotated} support functions.  To this end, for a distance of 10~\AA\ the six outer fullerenes were collectively rotated along the $z$-axis by angles of 15$^{\circ}$, 45$^{\circ}$ and 90$^{\circ}$, with the orientation of the central molecule remaining unchanged.  This was found to have a negligible impact on the IP and EA, with a difference in the values for the various orientations of around $0.01$~eV for the constrained fragment calculations compared with $0.05$~eV for the unconstrained cubic calculations.  Such values are too small to be significant compared to the errors associated with the basis.  In order to determine whether the energies are truly unaffected by the orientation, it would be necessary to account for the dispersion effects which are not captured by the LDA.  However, the fact that no spurious errors are introduced by the rotating of the support functions serves to further confirm the accuracy of the reformatting scheme.

\subsubsection{Comparison of environments}

\begin{figure*}
\centering
\subfigure[isolated fragment, $Q=-1$]{\includegraphics[width=0.24\textwidth]{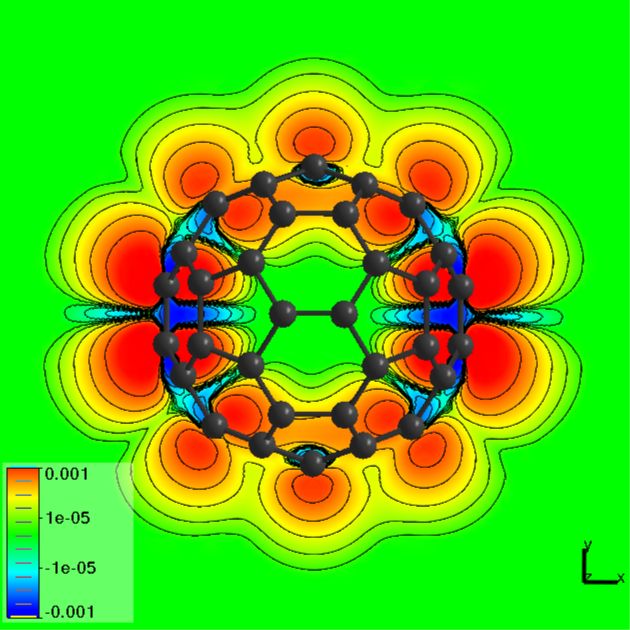}}
\subfigure[cubic, $Q=-1$]{\includegraphics[width=0.24\textwidth]{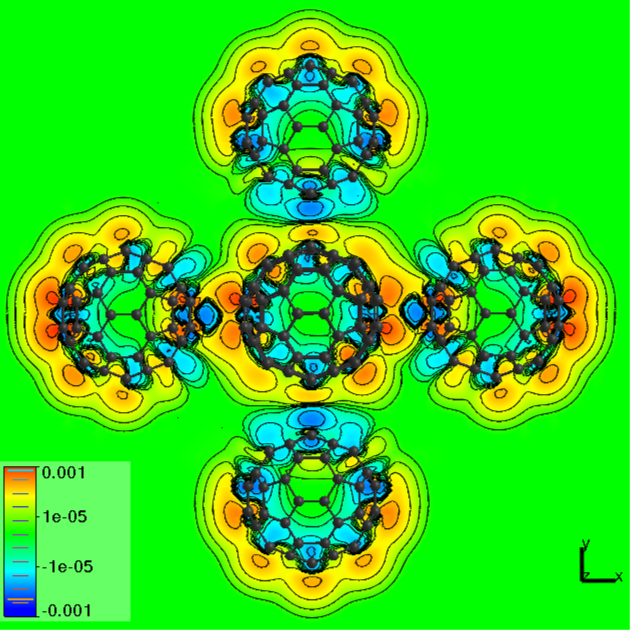}}
\subfigure[constrained fragment, $Q=-1$]{\includegraphics[width=0.24\textwidth]{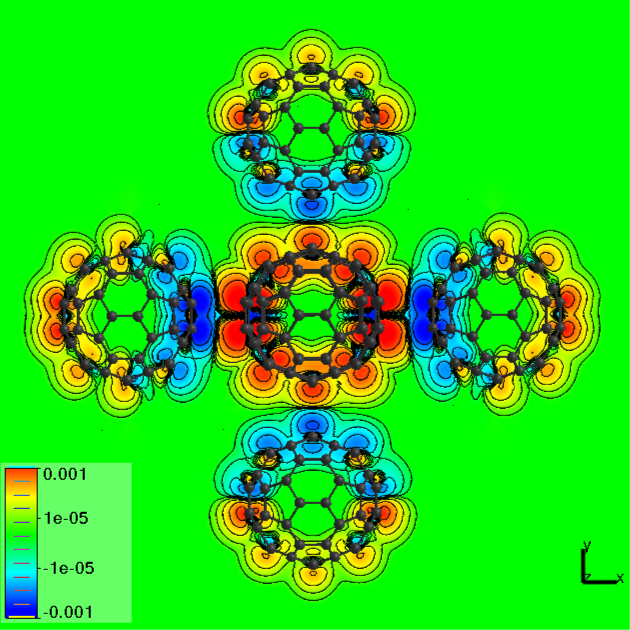}}
\subfigure[constrained fragment, $Q=-1$]{\includegraphics[width=0.24\textwidth]{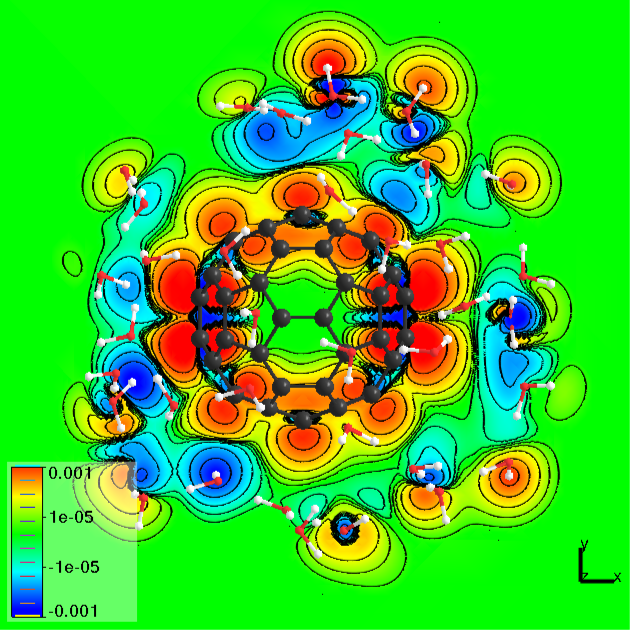}}
\subfigure[isolated fragment, $Q=+1$]{\includegraphics[width=0.24\textwidth]{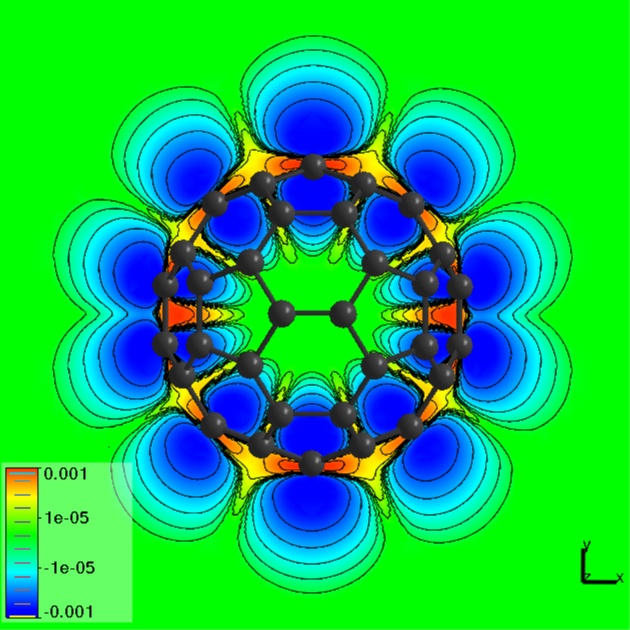}}
\subfigure[cubic, $Q=+1$]{\includegraphics[width=0.24\textwidth]{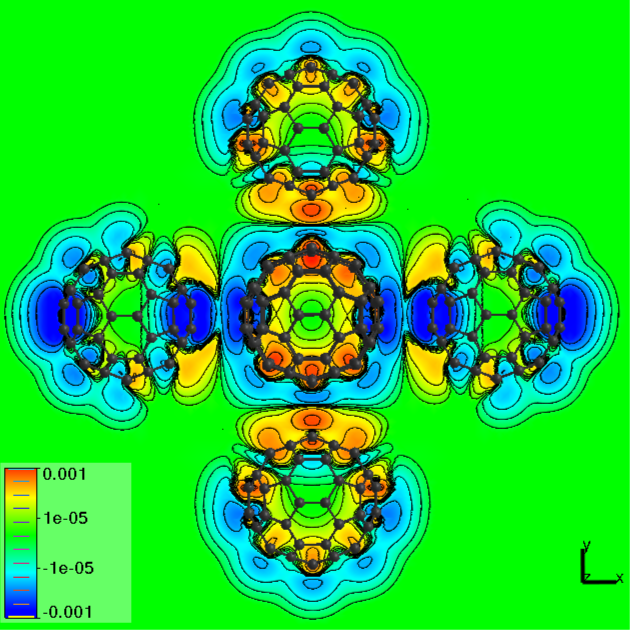}}
\subfigure[constrained fragment, $Q=+1$]{\includegraphics[width=0.24\textwidth]{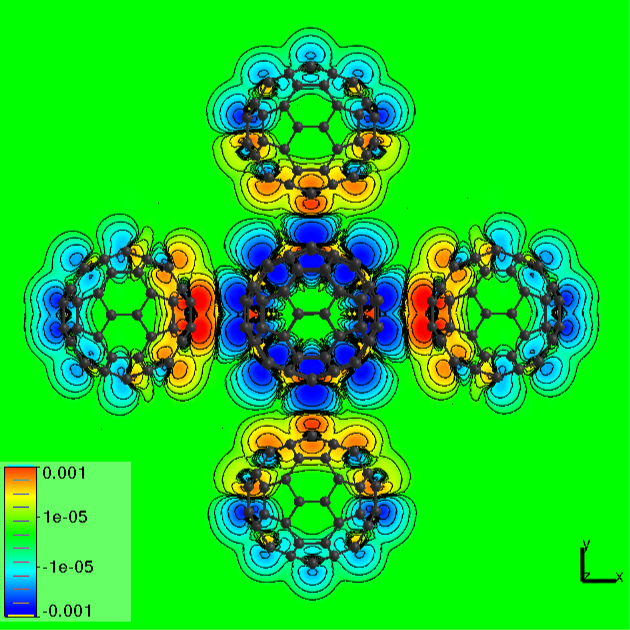}}
\subfigure[constrained fragment, $Q=+1$]{\includegraphics[width=0.24\textwidth]{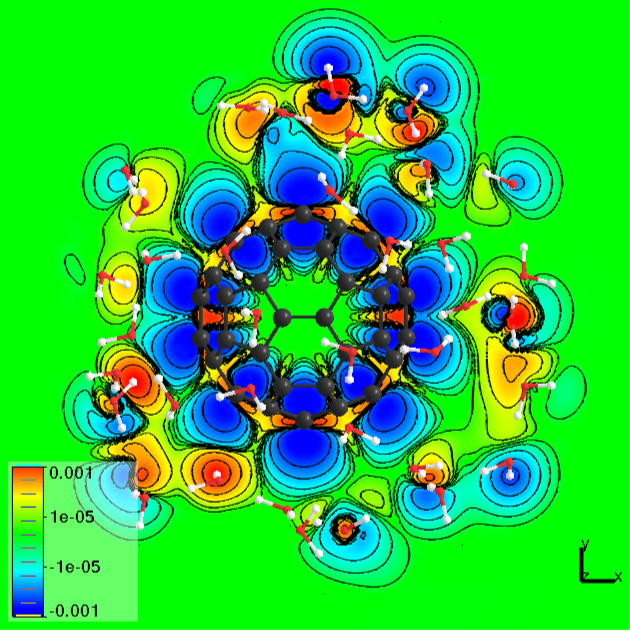}}
\caption{Density differences between the neutral and charged calculations for the fullerene when isolated, when surrounded by other fullerenes (with a center to center distance of 10~\AA) and when in water (structure A).  The densities are plotted on the central plane with a logscale, with red (blue) indicating an increase (decrease) in the electronic charge density with respect to the neutral.}\label{fig:densplots}
\end{figure*}

We now return to the results in Tab.~\ref{table:c60} in order to compare the effect of the two environments.
As expected, for both environments the constrained values
remain relatively close to the isolated results, certainly much closer than the unconstrained results for the fullerene environment.  In a sense, when the constraint is enforced,
the presence of the surrounding molecules could be thought of as a perturbation on the isolated state, although the
strength of the perturbation is clearly much stronger for the water.  To further explore this, we have also plotted differences in 
the converged electronic densities between the neutral and charged calculations in Fig.~\ref{fig:densplots}.
The effects of the charge constraint are clearly visible, with the excess charge distributed across the molecules
for the cubic calculation and much more clearly localized for the constrained calculations.
Furthermore, the charge difference on the central molecule for the constrained calculations
clearly retains the same character as the respective isolated density, with the excess or deficit of charge also
resulting in an induced dipole on the neighboring molecules.
As expected, the impact of the water is stronger than the neighboring fullerenes, where the closer proximity
and stronger dipole moment of the water molecules results in a stronger deviation from the isolated density
difference.  Similar behavior has also been observed for the water structures which are not depicted.

As can be seen from the variation in IP and EA for structures A, B and C (Tab.~\ref{table:c60}), 
the effect of the water
is not only stronger than that of the neighboring fullerenes, but the dependence on the structure
is also quite significant.  Furthermore, we have also observed that
the resulting energies are strongly affected by the choice of weight function, which is 
not the case for the fullerene environment.  There is some freedom in the procedure for
optimizing the template support functions, e.g.\ the localization radii and the number of additional states 
included; we tested a few of the different options.  For the fullerene environment the
variation in calculated IP and EA due to the choice of template parameters were small and systematic,
whereas for the aqueous environment
the variation was much stronger.  Indeed, the fragment approach provides an ideal setup
to explore the impact of different weight functions -- such a strong dependence would be harder
to detect when considering only two or three different choices.  In future the
choice of weight function could also be decoupled from the fragment basis to allow for a more
thorough exploration of its influence in constraining the charge.

Of course, in order to correctly assess the impact of the environment,
one should go beyond the model structures used here, both in terms of the size
and procedure used to generate them.  Furthermore, in the case of the water, proper sampling should be 
performed over a number of different configurations, which should be generated 
at the correct temperature e.g.\ using molecular 
dynamics~\cite{jcp/123/20/10.1063/1.2121647,jcp/138/4/10.1063/1.4789304} or 
Monte Carlo~\cite{Rivelino20062925,doi:10.1021/nl070308p} simulations.
However, aside from the generation of input structures and any eventual relaxations of the 
atomic coordinates, the fragment calculations are quick and easy to perform, 
requiring little additional setup aside from the template calculations. 
Furthermore, we have demonstrated both the accuracy and flexibility of the fragment approach
for such systems.
As such, given appropriate atomic coordinates this work could easily be extended in future to a large number of 
configurations for both environments, or indeed applied to other fullerenes or solvents.

\section{Conclusion}

We have presented a method for constrained DFT calculations on large systems, 
using a fragment based scheme.  This has been implemented in the BigDFT 
electronic structure code within a recently developed framework, which uses a 
basis of localized support functions represented in an underlying wavelet grid 
to achieve linear scaling behavior with respect to system size while retaining 
the systematic accuracy of the underlying grid.  The division of a given system 
into fragments (ideally distinct molecules), each with its own associated 
support functions leads to a natural approach to CDFT, %\luigi{Also this part  should go to the intro/abstract}
where the charge is constrained to a given fragment via a L\"owdin like 
definition of the weight function.  This L\"owdin approach can also be used to straightforwardly calculate
atomic charges, as we have demonstrated.

Furthermore, by using a 
reformatting scheme which enables the reuse of support functions for identical 
fragments, irrespective of their position or orientation in the system, we are 
able to further reduce the cost of simulations by an order of magnitude, as the 
support functions can be separately optimized for each `template' fragment and 
used as a fixed basis in the system of interest.  The properties of the wavelet basis set
ease the implementation of reformatting a 
numerical field given in real space, so that we were able to implement this scheme
in a manner which is both efficient and accurate.
The flexibility of this method, together with the ability of the BigDFT code to treat systems with many atoms (see e.g.\ Ref.~\cite{universal}), makes it ideally suited for %calculations on large systems.
%One could imagine for example optimizing the support functions for an isolated water molecule, and reusing these support functions in a system containing hundreds, or even thousands of water molecules (e.g.\ as a solvent), so that for the calculation performed on the large system one could keep the basis fixed and would therefore only need to optimize the density kernel.  
%Our approach therefore paves the way for 
both neutral and charged calculations on very large systems
at modest computational cost.

We have presented results from two previously studied systems in order to validate our approach, 
as well as an example application.  For this latter point, we have 
performed calculations on $\textrm{C}_{60}$ in two different environments, namely
a model nearest neighbor system containing seven fullerenes and in an 
aqueous solution.  The effects of the constraint
are clearly visible in the electronic densities, which we have compared to the unconstrained
and isolated results.  We have also shown that the presence of water has both a stronger impact on the results
and a stronger dependence on the choice of weight function than the presence of neighboring fullerenes.

%We have also investigated the effect of separation distance
%between $\textrm{C}_{60}$ molecules.

%We have also applied our formalism to the calculation of %on-site 
%energies and transfer integrals.
%where the reformatting scheme allows the sign to 
%be preserved, results for which will be presented elsewhere~\cite{mons_paper}.

The reformatting approach described here has another key benefit aside from reducing the cost 
of such calculations: 
as the basis set for each fragment  will 
remain equivalent following the reformatting, 
the computational setup provided by our approach is ideal where Hamiltonian matrix elements of the whole system have to be considered.
For example, for electronic coupling matrix elements (`transfer integrals') between two identical monomers, the basis set for each monomer will 
remain equivalent following the reformatting, so that there is no ambiguity in 
the sign of the coupling matrix elements.  In contrast, for support functions 
optimized from scratch there is no guarantee that the phase for both the support 
functions and wavefunction coefficients will be identical between the two 
molecules and thus the sign of the transfer integral cannot be determined.  
Indeed, we will be publishing results in the near future for such an application based on the framework presented in the current work~\cite{mons_paper}.

In the future we also hope to extend this work to permit calculations on realistic 
nanoscale devices, using a multi-scale approach.  In the first instance this 
would involve changing the definition of a fragment to an individual atom, which 
naturally leads to a DFT based tight-binding like method.  This formalism also 
allows the correct definition of an `embedded' approach, where different regions 
of a simulation cell are treated at different levels of precision, e.g.\ for a 
point defect in a bulk semiconductor, with higher accuracy close to the defect.
Work is ongoing in this direction.

\section*{Acknowledgements}

%is this correct?! copied and pasted from linear paper
We acknowledge funding from the European project MMM@HPC (RI-261594), the 
CEA-NANOSCIENCE BigPOL project and the ANR projects SAMSON (ANR-AA08-COSI-015) 
and NEWCASTLE (ANR-2010-COSI-005-01).
This research used resources of the Argonne Leadership Computing Facility at 
Argonne National Laboratory, which is supported by the Office of Science of the 
U.S. Department of Energy under contract DE-AC02-06CH11357.  CPU time was also 
provided by IDRIS (project i2014096905).

\bibliography{cdft}

\end{document}